\documentclass[11pt,letterpaper]{article}
\usepackage{soul}
\usepackage{jheppub}
\usepackage[utf8]{inputenc}
\usepackage[T1]{fontenc}
\usepackage{amsfonts}
\usepackage{amsmath}
\usepackage{graphicx}
\usepackage{subcaption}
\usepackage{tensor} 
\usepackage{mathtools} %
\allowdisplaybreaks[4]
\usepackage{amssymb}
\usepackage{amsthm,amsmath,amssymb}
\usepackage{mathrsfs}
\usepackage{euscript}
\usepackage{color}
\usepackage{braket}
\usepackage{orcidlink}

\newcommand{\rh}{r_{\text{h}}}
\newcommand{\dd}{\text{d}}

\title{ \boldmath Probing phase transitions of regular black holes in anti-de Sitter space with Lyapunov exponent}

\author[a,b]{Hao Xie\orcidlink{0009-0001-8207-2520}}
\emailAdd{xieh2017@lzu.edu.cn}

\author[a,b]{and Si-Jiang Yang\orcidlink{0000-0002-8179-9365}\footnote{Corresponding author}}
\emailAdd{yangsj@lzu.edu.cn}

\affiliation[a]{Lanzhou Center for Theoretical Physics, Key Laboratory of Theoretical Physics of Gansu Province, Key Laboratory of Quantum Theory and Applications of MoE, Gansu Provincial Research Center for Basic Disciplines of Quantum Physics, Lanzhou University, Lanzhou 730000, China\vspace{0.1cm}}

\affiliation[b]{ Institute of Theoretical Physics $\&$ Research Center of Gravitation, School of Physical Science and Technology, Lanzhou University, Lanzhou 730000, China \vspace{0.1cm}}

\abstract{We investigate the relationship between thermodynamic phase transitions and the Lyapunov exponent of charged regular anti-de Sitter black holes in quasi-topological gravity. Our results show that the Lyapunov exponent displays multivalued behavior during phase transitions. Moreover, along the coexistence curve the Lyapunov exponent changes discontinously and continuously at the critical point. Near the critical point, the Lyapunov exponent follows a power-law behavior with a critical exponent of $1/2$, suggesting its role as an order parameter and encodes information on black hole phase transitions.}


\begin{document}
\maketitle

\flushbottom

\section{Introduction} \label{Sec:introduction}

Black hole thermodynamics has long been a vibrant research frontier, driven by its deep and fundamental connections with thermodynamics, gravitation, and quantum field theory~\cite{Witten:2024upt,Padmanabhan:2009vy}. Black holes not only obey the four laws of thermodynamics~\cite{Bardeen:1973gs,Hawking:1971tu,Bekenstein:1972tm,Bekenstein:1974ax}, as do ordinary thermodynamic systems, but also possess an entropy~\cite{Bekenstein:1973ur} that can be interpreted as a Noether charge associated with the diffeomorphism invariance of the underlying theory~\cite{Iyer:1994ys, Wald:1993nt}. Black hole entropy implies the existence of an underlying microscopic structure~\cite{Strominger:1997eq,Wei:2015iwa,Maldacena:1996gb,Wei:2019uqg} and hints at a statistical origin for this entropy~\cite{Bekenstein:1975tw,York:1986it,Cheng:2024hxh,Cheng:2024efw,Liu:2025iei}. 

The thermodynamics of black holes in asymptotically AdS spacetimes has recently drawn considerable interest, owing to their rich phase structure and the profound implications of the anti-de Sitter/conformal field theory (AdS/CFT) correspondence~\cite{Maldacena:1997re,Aharony:1999ti,Witten:1998qj}. The Hawking–Page phase transition of a Schwarzschild AdS black hole~\cite{Hawking:1982dh}, describing the transition between black hole states and thermal radiation, can be interpreted in the dual gauge field theory as a confinement–deconfinement transition of the quark–gluon plasma~\cite{Witten:1998zw}. The small–large black hole phase transition in charged AdS black holes closely parallels the gas–liquid phase transition of van der Waals fluids~\cite{Chamblin:1999hg, Chamblin:1999tk}. Interpreting the negative cosmological constant as thermodynamic pressure~\cite{Kastor:2009wy}, known as extended phase space thermodynamics, makes the phase behavior of charged AdS black holes directly analogous to that of van der Waals systems~\cite{Kubiznak:2012wp}. Further studies uncovered rich phenomena in this framework~\cite{Cai:2013qga,Cheng:2016bpx,Xu:2022jyp,Xu:2024iji}, including triple points~\cite{Altamirano:2013uqa, Altamirano:2013ane, Wei:2014hba}, reentrant phase transitions~\cite{Frassino:2014pha}, $\lambda$ phase transitions~\cite{Hennigar:2016xwd,Bai:2023woh}, and isolated critical points~\cite{Dolan:2014vba,Hu:2024ldp,Ahmed:2022kyv,Yang:2025xck}, thereby highlighting deep parallels between black hole thermodynamics and conventional chemical thermodynamics.

In addition to being a thermodynamic system, a black hole is also a strong-gravity system that exhibits strong gravitational effects on spacetime and the evolution of matter, such as gravitational redshift, light bending, lensing, and gravitational wave emission. While strong gravity effects can be observed directly, the thermodynamic properties of black holes are far more elusive. This makes probing black hole thermodynamics through gravitational phenomena an especially intriguing research direction.

The Lyapunov exponent characterizes the exponential divergence or convergence of nearby trajectories in phase space, providing a fundamental measure of stability and chaos in physical systems~\cite{LYAPUNOV01031992,Hashimoto:2016dfz,Dalui:2018qqv,Zhang:2025cdx}. Within the AdS/CFT correspondence, the Lyapunov exponent of particles near a black hole horizon obeys a universal upper bound~\cite{Maldacena:2015waa}, though counterexamples exist~\cite{Lei:2023jqv,Lei:2024qpu,Dutta:2024rta}. It has been extensively employed to study the dynamics of unstable circular orbits in black hole spacetimes~\cite{Cardoso:2008bp, Fernando:2012ue, Sota:1995ms, Kan:2021blg}, and is further shown to be closely related to the imaginary part of certain quasinormal modes~\cite{Cardoso:2008bp}. Recent studies conjectured that the Lyapunov exponent of unstable circular orbits can serve as a probe of black hole thermodynamic phase structure~\cite{Guo:2022kio}. For black holes undergoing a first-order phase transition, the Lyapunov exponent exhibits multivalued behavior with respect to thermodynamic variables such as temperature and can serve as an order parameter for black hole phase transitions~\cite{Yang:2023hci,Lyu:2023sih,Kumara:2024obd}, a viewpoint further supported by subsequent systematic studies~\cite{Du:2024uhd,Shukla:2024tkw,Gogoi:2024akv,Chen:2025xqc,Yang:2025fvm,Ali:2025ooh,Kumar:2025kzt}.

Spacetime singularities, appearing both in gravitational collapse and at the beginning of the universe, are windows into physics beyond general relativity. When quantum effects are taken into account, it is widely accepted that spacetime singularities will disappear. One approach to resolving singularities is the construction of regular black hole metrics. Notable examples are Bardeen and Hayward black holes~\cite{bardeen1968non,Hayward:2005gi}, which arise as exact solutions of Einstein’s gravity coupled with nonlinear electrodynamics. However, the existence of such regular black hole metrics usually depends on exotic matter fields and delicate fine-tuning of coupling parameters and integration constants. As a result, they form only a measure-zero subset of the entire solution space of the field equations~\cite{Li:2024rbw}. Recently, it has been shown that by incorporating an infinite tower of higher-order curvature corrections into the Einstein–Hilbert action~\cite{Bueno:2024dgm}, quasi-topological gravity~\cite{Oliva:2010eb,Myers:2010ru,Dehghani:2011vu} offers a purely gravitational mechanism for resolving singularities in dimensions $D\geq 5$, leading to regular black hole solutions without the need for additional matter couplings~\cite{Bueno:2024zsx,Bueno:2024eig}.

Motivated by recent advances in the study of regular black holes within quasi-topological gravity, we explore the thermodynamics of charged regular black holes in this framework and examine their relationship with Lyapunov exponents. We study the evolution of the Lyapunov exponent across phase transitions under isothermal and isobaric conditions, and further examine the difference in Lyapunov exponents between small and large black holes along the coexistence line. Our results reveal that the Lyapunov exponent exhibits sudden changes across the first-order black hole phase transition and can serve as an effective indicator of this transition. Moreover, the difference between small and large black holes vanishes near the critical point, suggesting that it can serve as an order parameter.

The paper is organized as follows. In Sec.~\ref{sec:2}, we examine the thermodynamic phase transitions of charged regular AdS black holes in five- and seven-dimensional quasi-topological gravity. In Sec.~\ref{sec:3}, we investigate the Lyapunov exponents of null geodesics in these spacetimes and explore their connection to thermodynamic phase transitions. Finally, our main findings are summarized in Sec.~\ref{sec:con}.

\section{Regular black holes in quasi-topological gravity and its thermodynamics}\label{sec:2}

Recently, regular black holes in quasi-topological gravity have attracted considerable attention. Regular black hole solutions in quasi-topological gravity differ from the Bardeen~\cite{bardeen1968non,Ayon-Beato:1998hmi} and Hayward~\cite{Hayward:2005gi} black holes, as they can be obtained in a pure gravitational theory, independent of any matter field coupling. In this section, we present the charged regular AdS black holes in quasi-topological gravity and explore their thermodynamic phase transitions. 

\subsection{Charged regular AdS black holes in quasi-topological gravity}
In order to study small-large black hole phase transitions while maintaining the regularity of the solution, we consider quasi-topological gravity coupled to a Maxwell electromagnetic field. The corresponding action is given by~\cite{Bueno:2024zsx,Hennigar:2025ftm,Aguayo:2025xfi}
\begin{equation}
   \mathcal{I} = \mathcal{I}_{\text{QT}} + \mathcal{I}_{\text{EM}}, 
\end{equation}
where the gravitational part $\mathcal{I}_{\text{QT}}$ is given by
\begin{equation}
    \mathcal{I}_{\text{QT}} = \frac{1}{16\pi G} \int  \left( R -2 \Lambda + \sum_{n=2}^{\infty} \tilde{\alpha}_n \mathcal{Z}_n \right)\boldsymbol{\epsilon},
\end{equation}
and the Maxwell part takes the form
\begin{equation}
    \mathcal{I}_{\text{EM}} = - \frac{1}{16\pi G} \int F_{\mu \nu} F^{\mu \nu} \boldsymbol{\epsilon}.
\end{equation}
Here, $R$ denotes the Ricci scalar, $F_{\mu \nu}$ is the electromagnetic field strength tensor, $\Lambda$ represents the cosmological constant, and 
$\mathcal{Z}_n$ corresponds to the $n$-th order curvature term, which satisfies the following recursive relation~\cite{Bueno:2019ycr}
\begin{equation}
    \mathcal{Z}_{n+5} = - \frac{3(n+3)}{2(n+1)D(D-1)} \mathcal{Z}_1 Z_{n+4} + \frac{3(n+4)}{2nD(D-1)} \mathcal{Z}_2 \mathcal{Z}_{n+3} - \frac{(n+3)(n+4)}{2n(n+1)D(D-1)} \mathcal{Z}_3 \mathcal{Z}_{n+2}. \label{recurZ}
\end{equation}
Specifically, $\mathcal{Z}_1=R$ corresponds to the Einstein–Hilbert term, and the second-order quasi-topological density $\mathcal{Z}_2$ is a non-standard form of the Gauss–Bonnet term. Due to the considerable complexity of the full expressions for $\mathcal{Z}_3$, $\mathcal{Z}_4$ and $\mathcal{Z}_5$, the explicit expressions of $\mathcal{Z}_{1}-\mathcal{Z}_{5}$ are given in~\cite{Bueno:2024dgm,Bueno:2024eig,Hennigar:2025ftm}. Once the first five quasi-topological Lagrangian densities $\{\mathcal{Z}_i:i=1,...,5\} $ are determined, all higher-order quasi-topological Lagrangian densities can be systematically derived from Eq.~(\ref{recurZ}).

We consider a static, spherically symmetric metric:
\begin{equation}
    ds^2 = - N(r)^2 f(r) \dd t^2 + \frac{\dd r^2}{f(r)} + r^2 \Omega_{ij} \dd x^i \dd x^j,\label{metric}
\end{equation}
where $\Omega_{ij}$ represents the metric on the $(D-2)$-dimensional unit sphere $S^{D-2}$. By applying the reduced Lagrangian methods~\cite{Fels:2001rv,Deser:2003up}, the action of quasi-topological gravity, $\mathcal{I}_{\text{QT}}$, is evaluated using the metric ansatz from Eq.~(\ref{metric}). After performing integration by parts and discarding total derivative terms, the Lagrangian is simplified to
\begin{equation}
    \mathcal{I}_{\text{QT}} = \frac{(D-2)\Omega_{D-2}}{16 \pi G} \int dt dr N(r) \frac{d}{dr} \left[r^{D-1} h(\psi)\right],
\end{equation}
where
\begin{equation}
    h(\psi) = \frac{- 2 \Lambda}{(D-1)(D-2)}+ \psi + \sum_{n=2}^{\infty} \frac{(D-2n)}{D-2} \tilde{\alpha}_n \psi^n, \label{hpsi}
\end{equation}
and
\begin{equation}
    \psi \equiv \frac{1 - f(r)}{r^2}.
\end{equation}
The reduction of the Maxwell term can be performed in an analogous manner. With the static electric ansatz $A = \Phi(r) \dd t$, the Maxwell term in the action reduces to
\begin{equation}
    \mathcal{I}_{\text{EM}} = \frac{\Omega_{D-2}}{8 \pi G} \int dr \frac{r^{D-2} (\Phi'(r))^2}{N(r)}.
\end{equation}
By varying the reduced Lagrangian with respect to the unknown functions $N(r)$ and $f(r)$, one obtains
\begin{align}
            h'(\psi) N'(r) &= 0,\label{hN} \\
    \left[r^{D-1} h(\psi)\right]' &= -\frac{16\pi G}{(D-2) \Omega_{D-2}} \frac{\delta \mathcal{I}_\text{EM}}{\delta N}.\label{rhp}
\end{align}
From Eq.~\eqref{hN}, we find that $N$ is a constant. Therefore, without loss of generality, we set $N=1$. Varying the Maxwell part of the reduced action with respect to $\Phi(r)$, we can obtain
\begin{equation}
    \Phi(r) = \sqrt{\frac{D-2}{2(D-3)}}\frac{q}{r^{D-3}},
\end{equation}
where $q$ is an integration constant that is proportional to the electric charge. Integrating both sides of Eq.~\eqref{rhp} over $r$, we obtain
\begin{equation}
    h(\psi) =\frac{m}{r^{D-1}}-q^2 r^{4-2 D}, \label{Sr}
\end{equation}
where $m$ is an integration constant associated with the ADM mass $M$, given by
\begin{equation}
    M = \frac{m (D-2) \Omega_{D-2}}{16 \pi G}.
\end{equation}
When the coupling constants $\tilde{\alpha}_n$ are chosen as
\begin{equation}
    \tilde{\alpha}_n=\frac{\left[1-(-1)^n\right](D-2) \Gamma (\frac{n}{2})}{2\sqrt{\pi} (D-2n)\Gamma(\frac{n+1}{2})} \alpha^{n-1},
\end{equation}
using Eqs.~\eqref{hpsi} and \eqref{Sr}, the metric function can be obtained as~\cite{Hennigar:2025ftm}
\begin{equation}
    f(r) = 1-\frac{r^2 \mathcal{S}(r)}{\sqrt{1+{\alpha}^2 \mathcal{S}^2 (r)}}, \label{fr}
\end{equation}
where $\mathcal{S}(r)$ is defined as
\begin{equation}
\begin{aligned}
        \mathcal{S}(r) &= h(\psi) + \frac{ 2 \Lambda}{(D-1)(D-2)} \\
             &= \frac{ 2 \Lambda}{(D-1)(D-2)}+\frac{m}{r^{D-1}}-q^2 r^{4-2 D}. 
\end{aligned}
\end{equation}

The metric given in Eq.~\eqref{metric}, with the metric function specified in Eq.~\eqref{fr}, describes a black hole whose horizon radius $\rh$ is determined by the condition $f(\rh)=0$.
To investigate the regularity of this spacetime, it is necessary to examine the behavior of $f(r)$ as $r \rightarrow 0$,
\begin{equation}
    f(r) = 1+\frac{r^2}{|\alpha|} + \mathcal{O}(r^3).\label{limr0}
\end{equation}
Based on Eq.~(\ref{limr0}), in the vicinity of $r=0$, the charged regular AdS black hole solution given by Eq.~(\ref{metric}) in quasi-topological gravity smoothly reduces to Minkowski spacetime. Consequently, both massive and massless particles moving along geodesics do not encounter geodesic incompleteness or termination at $r=0$; instead, their trajectories can be smoothly extended across this point. As a result, all timelike and null geodesics are complete, indicating the regularity of the spacetime. Furthermore, the Ricci scalar and the Kretschmann scalar remain finite at $r=0$~\cite{Hennigar:2025ftm}, further supporting the nonsingular nature of the spacetime.

It is worth emphasizing that the regularity of the solution does not rely on the Maxwell field. When the electromagnetic field is switched off, the charge parameter $q$ in the function $\mathcal{S}(r)$ vanishes. In this case, the metric function $f(r)$ exhibits the following asymptotic behavior as $r \rightarrow 0$:
\begin{equation}
    f(r) = 1-\frac{r^2}{|\alpha|} + \mathcal{O}(r^3).\label{limr00}
\end{equation}
The sign difference between Eqs.~\eqref{limr0} and~\eqref{limr00} originates from the fact that, in the presence of an electromagnetic field, the charge-dependent term dominates the near-origin behavior of the metric function. A detailed analysis further demonstrates that the geodesics remain complete, and both the Ricci scalar and the Kretschmann scalar are finite throughout the entire spacetime~\cite{Hennigar:2025ftm}. The results show that the regularity of the spacetime in quasi-topological gravity is independent of the Maxwell field.

\subsection{Thermodynamics and phase transitions of charged regular AdS black holes} \label{qstherom}

In the extended phase space, the negative cosmological constant is interpreted as the thermodynamic pressure, while the black hole mass is regarded as the enthalpy rather than the internal energy~\cite{Kastor:2009wy}. Within this framework, the first law of black hole thermodynamics and the corresponding Smarr relation for charged regular black holes in quasi-topological gravity can be expressed as~\cite{Hennigar:2025ftm}:
\begin{align}
        dM &= TdS + \Phi_{\text{EM}} d Q + VdP + \Psi d \alpha,\\
    (D-3)M &= (D-2)TS + (D-3) \Phi_{\text{EM}} Q - 2VP + 2\Psi \alpha.
\end{align}
Here, the coupling constant $\alpha$ is treated as a thermodynamic variable. In these relations, $T$ denotes the Hawking temperature, $S$ is the Wald entropy, $Q$ represents the electric charge, and $\Phi_{\text{EM}}$ is its conjugate electric potential. The quantity $P$ corresponds to the thermodynamic pressure in the extended phase space, with $V$ being its conjugate thermodynamic volume. The term $\Psi$ denotes the potential conjugate to the coupling constant $\alpha$.
The explicit expressions of these thermodynamic variables are given as follows~\cite{Hennigar:2025ftm}
\begin{align}
    M &= \frac{(D-2) \Omega_{D-2}}{16 \pi G} \rh ^{D-1} (\frac{1}{\sqrt{\rh^4-\alpha ^2}}+\frac{1}{l^2}+\frac{q^2}{\rh^{2D-4}} ),\\
     T &= -\frac{\left(\rh^4-\alpha ^2\right)  \left[(D-3) l^2 q^2 \rh^{-2D} \sqrt{\rh^4-\alpha ^2}-(D-1) \rh^{-4} \left(\sqrt{\rh^4-\alpha ^2}+l^2\right)\right]+2 l^2}{4 \pi  l^2 \rh}, \label{temp}\\
    S &= \frac{\Omega _{D-2} \rh^{D-2} \, }{4 G}  \,_2F_1\left(\frac{3}{2},\frac{1}{2}-\frac{D}{4};\frac{3}{2}-\frac{D}{4};\frac{\alpha ^2}{\rh^4}\right),\label{entropy}\\
    Q &= \frac{\sqrt{2(D-2)(D-3)}}{8 \pi G} \Omega_{D-2} q,\\
    \Phi_{\text{EM}} &= \sqrt{\frac{D-2}{2(D-3)}}\frac{q}{r_{\text{h}}^{D-3}},\\
    V &= \frac{\Omega_{D-2}\rh^{D-1}}{D-1},\\
    P &= \frac{(D-1)(D-2)}{16 \pi G l^2} \label{press},
\end{align}
where $l$ denotes the AdS radius, which is related to the cosmological constant by
\begin{equation}
    \Lambda=-\frac{(D-1)(D-2)}{2l^2},
\end{equation}
and $_2F_1(a,b;c;z)$ is the hypergeometric function.

In order to obtain the equation of state for the black holes, we introduce the specific volume and the molecular volume parameter associated with the black holes.~\cite{Hennigar:2025ftm}
\begin{equation}
    \begin{split}
        v = \frac{4G}{D-2} \rh, \quad b = \frac{4G\sqrt{\alpha}}{D-2}. \label{vbdefin}
    \end{split}
\end{equation}
We set $G=1$ for convenience in subsequent discussions. Using the black hole temperature in Eq.~(\ref{temp}), one can directly derive the equations of state for charged regular AdS black holes in five and seven dimensions within quasi-topological gravity, which are given by:
\begin{itemize}
    \item Five-dimensional case:
\begin{equation}
     P=\frac{T v^5}{\left(v^4-b^4\right)^{3/2}}-\frac{2 \left(v^4-2 b^4\right)}{3 \pi  \left(v^4-b^4\right)^{3/2}}+\frac{512 q^2}{243 \pi  v^6}; \label{eos5D}
\end{equation}
\item Seven-dimensional case:
\begin{equation}
     P=\frac{T v^5}{\left(v^4-b^4\right)^{3/2}}+\frac{2 \left(3 b^4-2 v^4\right)}{5 \pi  \left(v^4-b^4\right)^{3/2}}+\frac{262144 q^2}{1953125 \pi  v^{10}}. \label{eos7D}
\end{equation}
\end{itemize}

To investigate the thermodynamic stability of black holes and to further explore their phase structure, we consider the Gibbs free energy, defined by
\begin{equation}
    F = M - TS.
\end{equation}
For the five- and seven-dimensional cases, the Gibbs free energy takes the following forms, respectively:
\begin{itemize}
    \item Five-dimensional case:
\begin{equation}
    \begin{aligned}
        F=\frac{9 \pi  q^2}{16 \rh^2}+\frac{3 \pi  \rh^8}{16 \left(\rh^4-\alpha ^2\right){}^{3/2}}+\frac{3 \pi ^2 T \rh^9}{8 \left(\rh^4-\alpha ^2\right){}^{3/2}}-\frac{1}{2} \pi ^2 \rh^3 T \, _2F_1\left(-\frac{3}{4},\frac{3}{2};\frac{1}{4};\frac{\alpha ^2}{\rh^4}\right);
\end{aligned}
\end{equation}
\item Seven-dimensional case:
\begin{equation}
    \begin{aligned}
        F=\frac{25 \pi ^2 q^2}{48 \rh^4}+\frac{5 \pi ^2 \rh^{10}}{48 \left(\rh^4-\alpha ^2\right)^{3/2}}+\frac{5 \pi ^3 \rh^{11} T}{24 \left(\rh^4-\alpha ^2\right)^{3/2}}-\frac{1}{4} \pi ^3 \rh^5 T \, _2F_1\left(-\frac{5}{4},\frac{3}{2};-\frac{1}{4};\frac{\alpha ^2}{\rh^4}\right).
\end{aligned}
\end{equation}
\end{itemize}

\begin{figure}[ht]
	\begin{subfigure}{.5\textwidth}
		\centering
		\includegraphics[width=.93\linewidth]{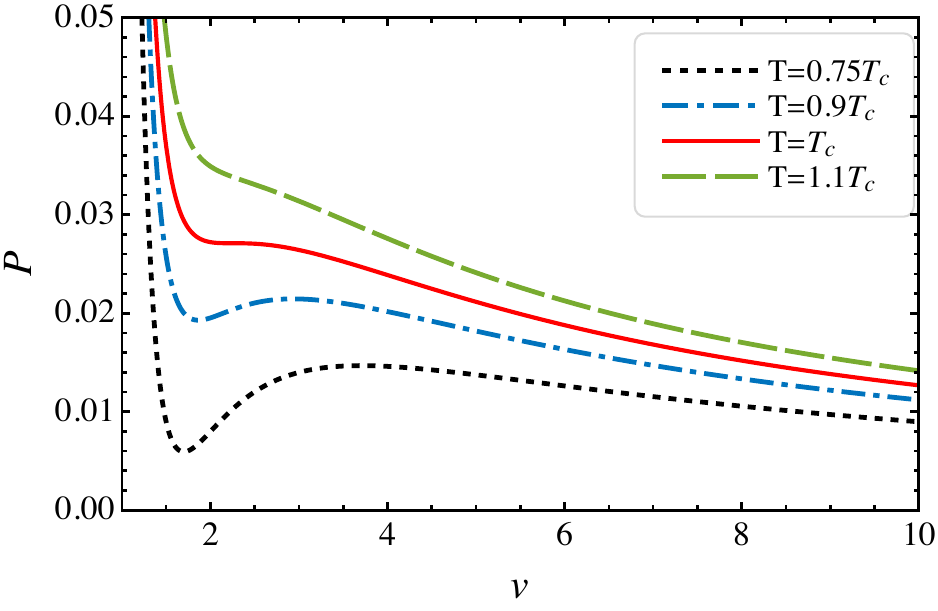}  
		\caption{}
		\label{fig:5DPv}
	\end{subfigure}
\begin{subfigure}{.5\textwidth}
	\centering
	\includegraphics[width=.9\linewidth]{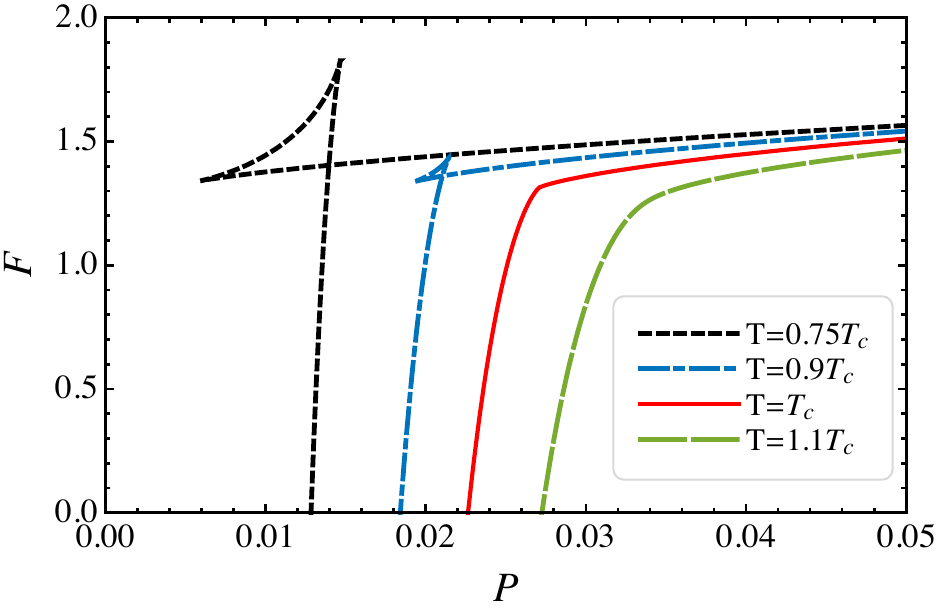}  
	\caption{}
	\label{fig:5DFP}
\end{subfigure}
\caption{(a). The van der Waals–like behavior in the $P-v$ plane for five-dimensional charged regular AdS black holes in quasi-topological gravity; (b). The swallowtail structure in the $F-P$ plane for five-dimensional charged regular AdS black holes in quasi-topological gravity. Parameters are chosen as $q=1/2$, $b=4/5~(\alpha = 9/25)$.}
\label{fig:5DPcFP}
\end{figure}

\begin{figure}[ht]
	\begin{subfigure}{.5\textwidth}
		\centering
		\includegraphics[width=.93\linewidth]{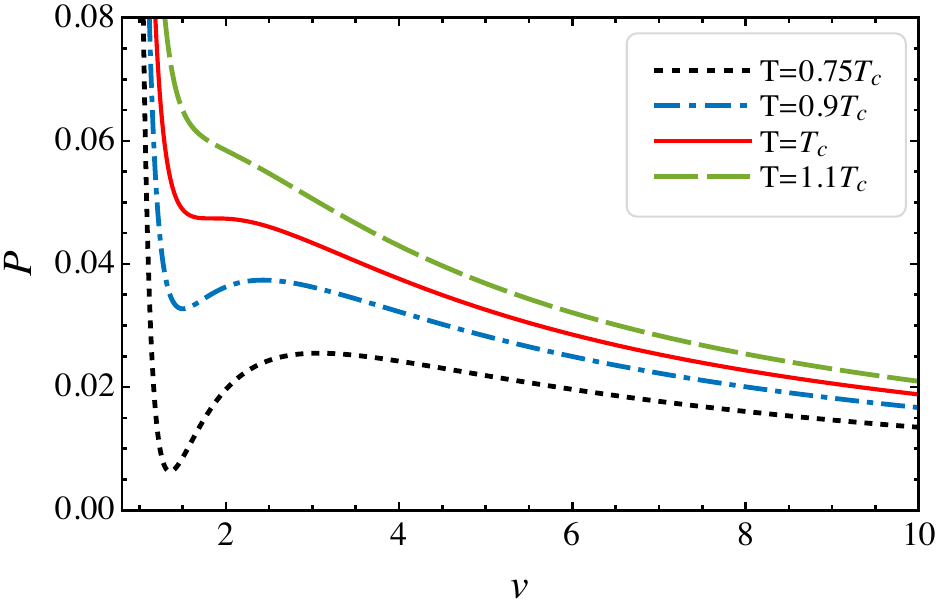}  
		\caption{}
		\label{fig:7DPv}
	\end{subfigure}
\begin{subfigure}{.5\textwidth}
	\centering
	\includegraphics[width=.94\linewidth]{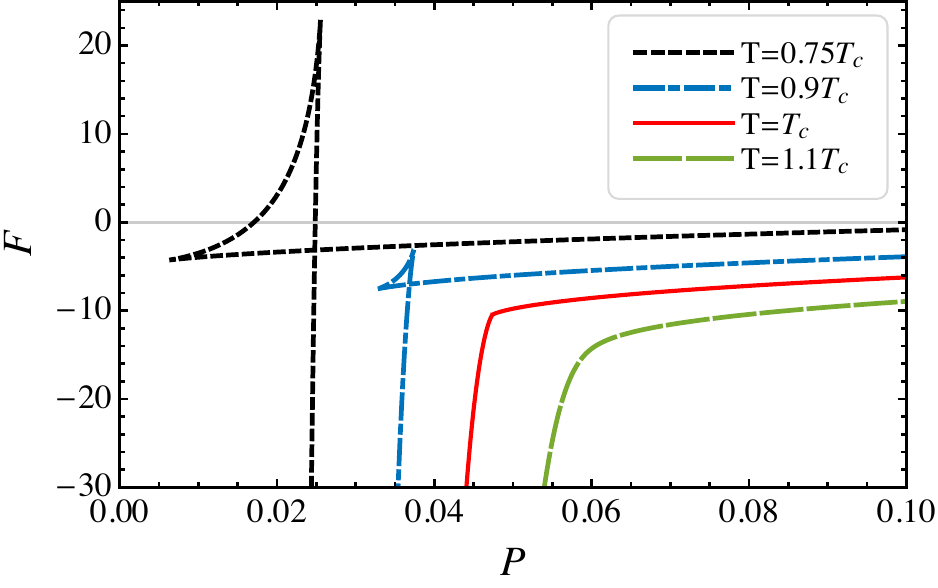}  
	\caption{}
	\label{fig:7DFP}
\end{subfigure}
\caption{(a). The van der Waals–like behavior in the $P-v$ plane for seven-dimensional charged regular AdS black holes in quasi-topological gravity; (b). The swallowtail structure in the $F-P$ plane for seven-dimensional charged regular AdS black holes in quasi-topological gravity. Parameters are chosen as $q=1/2$, $b=4/5~(\alpha = 1)$.}
\label{fig:7DPvFP}
\end{figure}

\begin{figure}[ht]
	\begin{subfigure}{.5\textwidth}
		\centering
		\includegraphics[width=.93\linewidth]{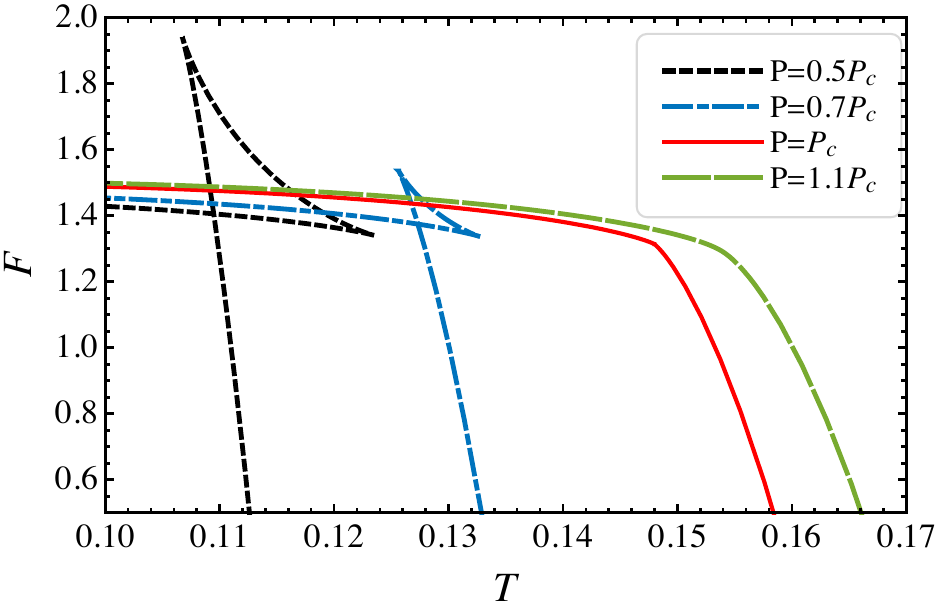}  
		\caption{}
		\label{fig:5DFT}
	\end{subfigure}
\begin{subfigure}{.5\textwidth}
	\centering
	\includegraphics[width=.9\linewidth]{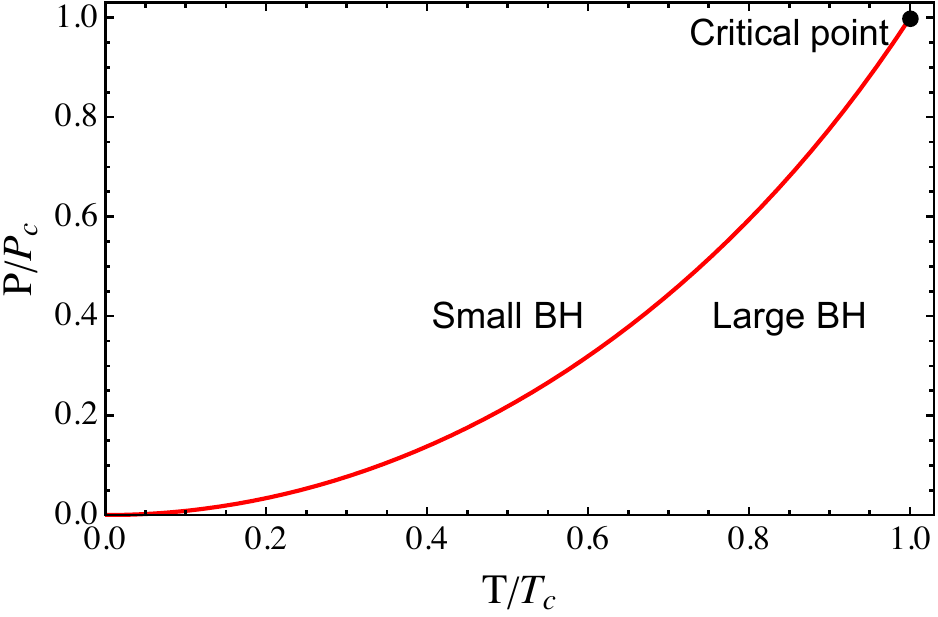}  
	\caption{}
	\label{fig:5DCoe}
\end{subfigure}
\caption{(a). The swallowtail structure in the $F-T$ plane for five-dimensional charged regular AdS black holes in quasi-topological gravity. (b). The coexistence curve of phase transition for the five-dimensional charged regular AdS black holes in quasi-topological gravity. Parameters are chosen as $q=1/2$, $b=4/5~(\alpha = 9/25)$.}
\label{fig:5DCoeFT}
\end{figure}

\begin{figure}[ht]
	\begin{subfigure}{.5\textwidth}
		\centering
		\includegraphics[width=.96\linewidth]{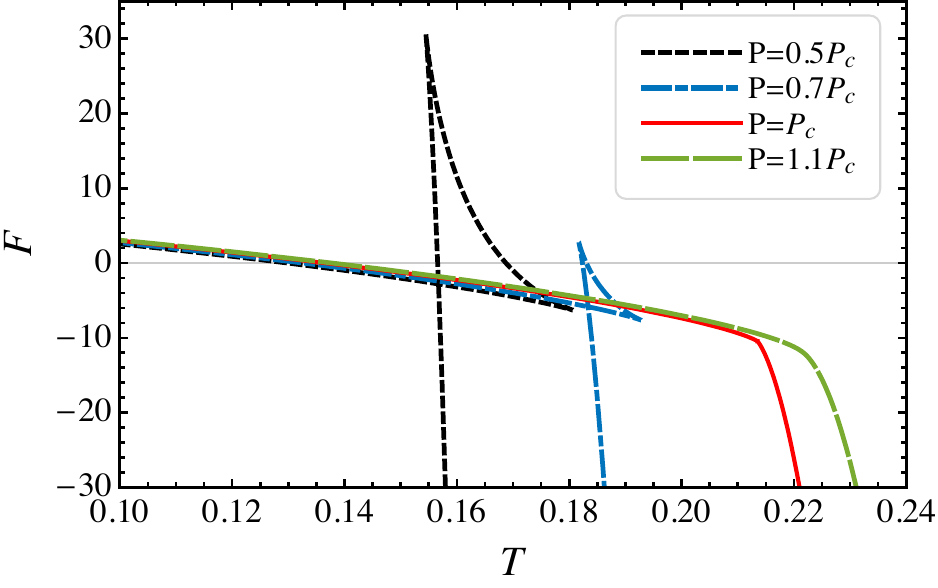}  
		\caption{}
		\label{fig:7DFT}
	\end{subfigure}
\begin{subfigure}{.5\textwidth}
	\centering
	\includegraphics[width=.93\linewidth]{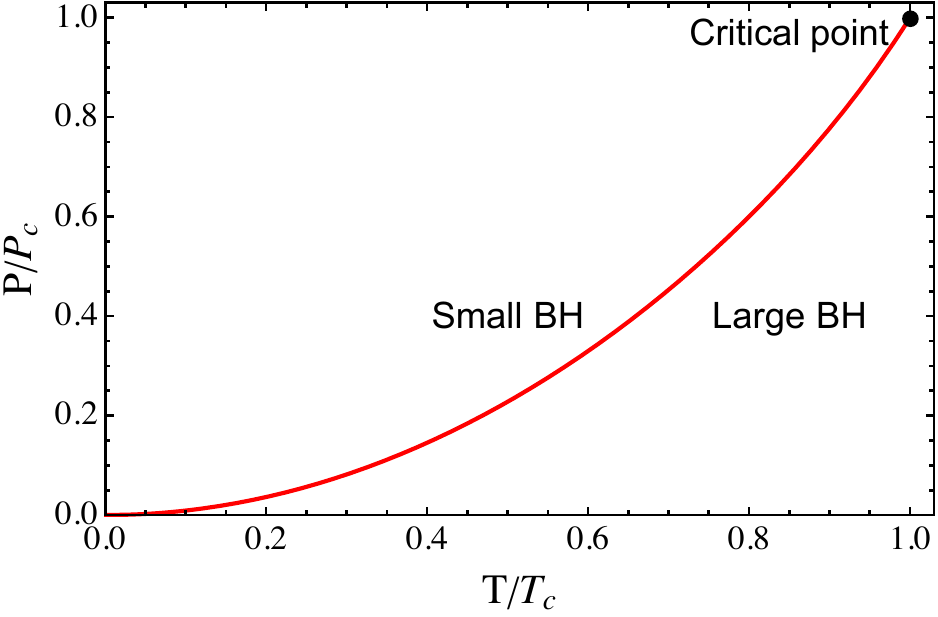}  
	\caption{}
	\label{fig:7DCoe}
\end{subfigure}
\caption{(a). The swallowtail structure in the $F-T$ plane for seven-dimensional charged regular AdS black holes in quasi-topological gravity. (b). The coexistence curve of phase transition for the seven-dimensional charged regular AdS black holes in quasi-topological gravity. Parameters are chosen as $q=1/2$, $b=4/5~(\alpha = 1)$.}
\label{fig:7DCoeFT}
\end{figure}

To illustrate the thermodynamic behavior and phase transitions, the $P-v$ and $F-P$ curves of the five- and seven-dimensional black holes at different temperatures are presented in Figs.~\ref{fig:5DPcFP} and~\ref{fig:7DPvFP}, respectively. As shown in Figs.~\ref{fig:5DPv} and~\ref{fig:7DPv}, when the temperature is below the critical temperature ($T<T_{\text{c}}$), the $P-v$ curves exhibit van der Waals–like behavior, while the corresponding free energy diagrams in Figs.~\ref{fig:5DFP} and~\ref{fig:7DFP} display the characteristic swallowtail structure, signaling a first-order phase transition between small and large black holes. At the critical temperature ($T=T_{\text{c}}$), indicated by the red curves, both the van der Waals–like behavior in the $P-v$ plots and the swallowtail structure in the $F-P$ plots disappear, suggesting a second-order phase transition at the critical point ($T=T_{\text{c}}, P=P_{\text{c}}$). For $T>T_{\text{c}}$, these features disappear completely, indicating the absence of any phase transition.

Physical quantities along the coexistence curve typically display rich and nontrivial behaviors~\cite{Wei:2023mxw}. To facilitate the subsequent analysis of the Lyapunov exponent between the small and large black hole phase transition, we begin by investigating the coexistence curve. The coexistence curve corresponding to the phase transition is obtained by imposing the condition that the Gibbs free energies and temperatures of the small and large black hole phases are the same. Using polynomial fitting techniques~\cite{Wei:2014qwa}, we obtain the coexistence curves for five- and seven-dimensional charged regular AdS black holes, as shown in Figs.~\ref{fig:5DCoe} and \ref{fig:7DCoe}, respectively. Each point in the coexistence curves corresponds to a thermodynamic equilibrium state where small and large black holes coexist at the same temperature and pressure. For isobaric or isothermal thermodynamic processes occurring below the critical point, crossing the coexistence curve induces a small-large black hole phase transition. Such a transition is analogous to the liquid-gas phase transition in a van der Waals fluid and is characterized by the van der Waals–like behavior in the $P-v$ diagrams and the swallowtail structure in the free energy diagrams, as shown in Figs.~\ref{fig:5DPcFP},~\ref{fig:7DPvFP},~\ref{fig:5DFT} and~\ref{fig:7DFT}. Conversely, above the critical point, the black hole remains in a single thermodynamic phase and does not exhibit any phase transition. 
It should be emphasized that the critical ratios of thermodynamic variables for charged regular AdS black holes in quasi-topological gravity are dependent on the parameters 
$q$, $\alpha$, and the spacetime dimension. Such dependence is in clear contrast to the universal behavior observed in van der Waals fluids and Reissner–Nordstr\"om AdS black holes.

\section{Lyapunov exponent and black hole phase transitions}\label{sec:3}

As emphasized earlier, a black hole is not only a thermodynamic system but also a strong gravitational system. In this section, we investigate the motion of test particles in the spacetime of charged regular AdS black holes within the framework of quasi-topological gravity, and investigate how the behavior of timelike and null geodesics reflects the underlying thermodynamic characteristics of the black holes.

\subsection{Unstable circular orbits}

The motion of test particles in a static, spherically symmetric black hole spacetime is governed by the geodesic equation, which can be derived from the Lagrangian. In five-dimensional spacetime, the Lagrangian for the motion of test particles takes the form
\begin{equation}
\begin{split}
    \mathcal{L} &=\frac{1}{2}g_{\mu\nu}\dot{x}^\mu\dot{x}^\nu\\
    &= \frac{1}{2} \left[-f(r)\dot{t}^2 + \frac{1}{f(r)}\dot{r}^2 + r^2(\dot{\theta_1}^2 + \sin^2 \theta_1 \dot{\theta}_2 ^2 + \sin^2 \theta_1 \sin^2 \theta_2 \dot{\varphi}^2) \right],
\end{split}
\end{equation}
where we have denoted $dx^\mu/d\lambda$ as $\dot{x}^\mu$, with $\lambda$ as the affine parameter. Similarly, in seven-dimensional spacetime, the Lagrangian for the motion of test particles takes the form
\begin{equation}
  \begin{aligned}
     \mathcal{L} = \frac{1}{2} \bigg[ -f(r) \dot{t}^2 + \frac{1}{f(r)} \dot{r}^2 + r^2 \big(
\dot{\theta}_1^2 + \sin^2 \theta_1 \, \dot{\theta}_2^2 + \sin^2 \theta_1 \sin^2 \theta_2 \, \dot{\theta}_3^2 \\
+ \sin^2 \theta_1 \sin^2 \theta_2 \sin^2 \theta_3 \, \dot{\theta}_4^2 + \sin^2 \theta_1 \sin^2 \theta_2 \sin^2 \theta_3 \sin^2 \theta_4 \, \dot{\varphi}^2
\big) \bigg].
  \end{aligned}
\end{equation}
The spherical symmetry of the spacetime allows for a simplification of the particles' orbital motion. Specifically, this symmetry enables us to restrict the particles' orbit to a sub-manifold, where all angular coordinates except $\varphi$ are fixed at the constant value $\pi/2$. Under this convention, the Lagrangian of the particles is given by:
\begin{equation}
    \mathcal{L} = \frac{1}{2} \left[-f(r) {\dot{t}}^2 + \frac{1}{f(r)} {\dot{r}}^2 + r^2 {\dot{\varphi}}^2\right]. \label{particleL}
\end{equation}
Since the spacetime is static and spherically symmetric, the energy $E$ and angular momentum $L$ of particles in the spacetime are conserved.
The energy and the angular momentum of particles are given by
\begin{equation}
    \begin{split}
        E = -f(r) \dot{t}, \qquad L = r^2 \dot{\varphi}.\label{noetherEL}
    \end{split}
\end{equation}
Using Eq.~(\ref{noetherEL}), the equation of radial motion can be expressed as
\begin{equation}
    \dot{r}^2 + V_{\text{eff}} (r) = E^2,
\end{equation}
where $V_{\text{eff}}$ denotes the effective potential, which is explicitly given by
\begin{equation}
    V_{\text{eff}} (r) = f(r)(\frac{L^2}{r^2} + \delta),\label{Veffm}
\end{equation}
with $\delta = 0$ for massless particles and $\delta = 1$ for massive particles. The unstable circular orbits satisfy the following conditions:
\begin{equation}
   \begin{split}
           V'_{\text{eff}}(r_{\text{c}})=0, \quad V''_{\text{eff}}(r_{\text{c}})<0. \label{Veff}
   \end{split}
\end{equation}
where $r_{\text{c}}$ is the radius of the unstable circular orbits.

As indicated by Eq.~\eqref{Veffm}, the effective potentials of massive and massless particles exhibit distinct forms, which in turn give rise to different existence criteria and radial locations for their unstable circular orbits. In what follows, we determine the critical conditions of these orbits through a detailed analysis of the respective effective potentials.

For massless and massive particles, the circular orbits are determined, respectively, by
\begin{equation}
    2 f(r_\text{c}) - r_\text{c}f'(r_\text{c}) = 0,\label{cgeom0}
\end{equation}
\begin{equation}
    L^2 = \frac{r_\text{c}^3 f'(r_\text{c})}{2f(r_\text{c})-r_\text{c}f'(r_\text{c})},\label{cgeom1}
\end{equation}
where prime denotes derivative with respect to coordinate $r$. As shown in Eq.~\eqref{cgeom0}, the circular orbits of massless particles are determined by the spacetime geometry and are independent of the particles’ parameters, whereas those of massive particles, as given by Eq.~\eqref{cgeom1}, depend explicitly on their angular momentum. Using Eq.~(\ref{cgeom1}), we find that in the large black hole phase at temperature $T = 0.75 T_{\text{c}}$, unstable circular orbits exist only when the angular momentum exceeds certain threshold values. 
For charged regular AdS black holes with $q=1/2$ and $b=4/5$, the corresponding conditions for the existence of unstable circular orbits are given as follows:
\begin{itemize}
    \item Five-dimensional case:
    \begin{equation}
        L  > 29.6867; \label{5DLcon}
    \end{equation}
    \item Seven-dimensional case:
    \begin{equation}
         L  > 31.9241. \label{7DLcon}
    \end{equation}
\end{itemize}

\begin{figure}[ht]
	\begin{subfigure}{.5\textwidth}
		\centering
		\includegraphics[width=.93\linewidth]{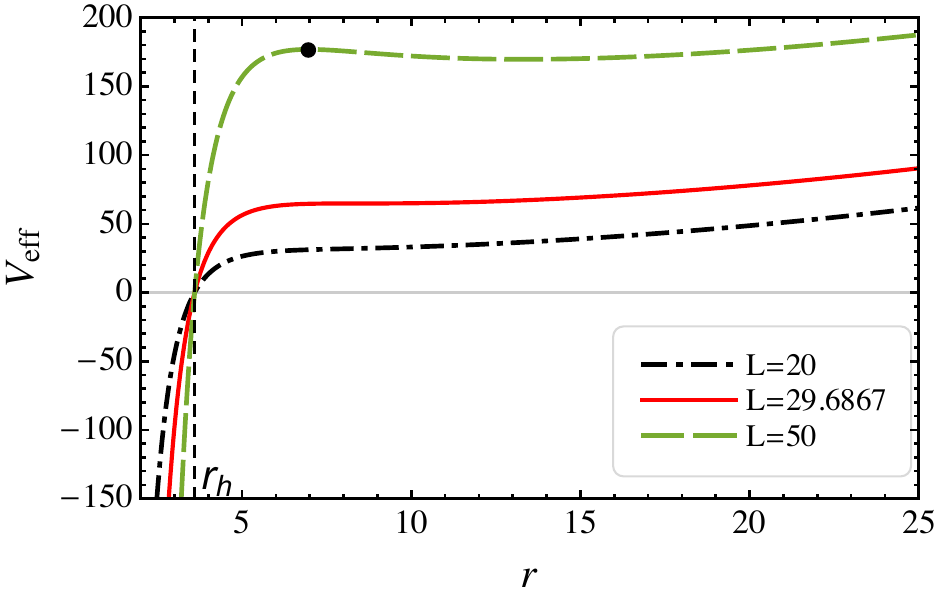}  
		\caption{}
		\label{fig:5DLcmass}
	\end{subfigure}
\begin{subfigure}{.5\textwidth}
	\centering
	\includegraphics[width=.93\linewidth]{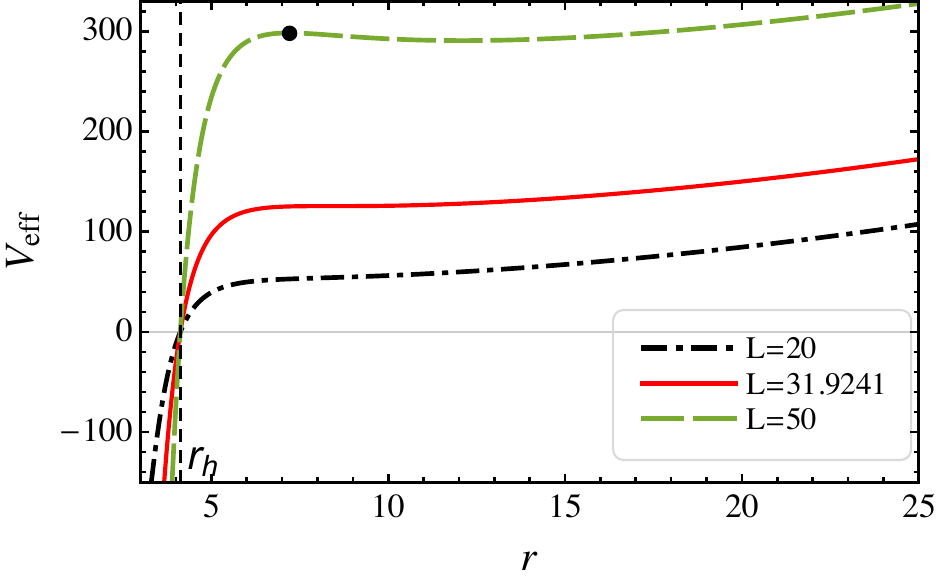}  
	\caption{}
	\label{fig:7DLcmass}
\end{subfigure}
\caption{Effective potential $V_{\text{eff}}$ of massive particles with varying angular momentum $L$ as a function of the radial coordinate $r$ in the large black hole phase at $T = 0.75 T_{\text{c}}$. The parameters are set to $q=1/2$ and $b=4/5$.
(a) five-dimensional case; (b) seven-dimensional case. }
\label{fig:5D7DLcmass}
\end{figure}

In Fig.~\ref{fig:5D7DLcmass}, we present the effective potential of massive particles with different angular momenta $L$ in the large black hole phase of five- and seven-dimensional charged regular AdS black holes. As shown in Figs.~\ref{fig:5DLcmass} and \ref{fig:7DLcmass}, when $L$ satisfies Eqs.~(\ref{5DLcon}) and~(\ref{7DLcon}), the effective potential exhibits a local maximum, corresponding to an unstable circular orbit. Conversely, if the angular momentum $L$ does not satisfy these conditions, the effective potential $V_{\text{eff}}$ increases monotonically with the radial coordinate $r$, indicating the absence of any unstable circular orbit.

From Eq.~(\ref{cgeom0}) and Eq.~(\ref{cgeom1}), we observe that the unstable circular orbits of massless particles emerge in the limit where the angular momentum of massive particles tends to infinity. Consequently, unstable circular orbits for massless particles exist in all cases, and their radii are always smaller than those of massive particles.

\subsection{Lyapunov exponent and black hole phase transitions}

The Lyapunov exponent is a key measure of a dynamical system’s sensitivity to initial conditions, quantifying the average exponential rate at which nearby trajectories diverge or converge over time~\cite{Hashimoto:2016dfz}. In this subsection, we introduce the Lyapunov exponents for both massless and massive particles and explore their connection with the phase transitions of charged regular AdS black holes in quasi-topological gravity.

For test particles in circular orbits, the Lyapunov exponents, which characterize orbital instability, are given by:
\begin{itemize}
    \item Massless particles:
    \begin{equation}
        \lambda = \sqrt{-\frac{f(r_{\text{c}}) r_{\text{c}} ^2}{2 L^2} V''_{\text{eff}}(r_{\text{c}})};\label{mass0}
    \end{equation}
    \item Massive particles:
    \begin{equation}
        \lambda = \frac{1}{2} \sqrt{\left[r_{\text{c}} f'(r_{\text{c}}) - 2 f(r_{\text{c}})\right] V''_{\text{eff}} (r_{\text{c}})}.\label{mass1}
    \end{equation}
\end{itemize}
These Lyapunov exponents provide a quantitative characterization of orbital instability: a larger $\lambda$ indicates a faster divergence of nearby trajectories and thus a more unstable orbit. A detailed derivation of Eqs.~\eqref{mass0} and~\eqref{mass1} is presented in Appendix~\ref{appendixA}.

\begin{figure}[ht]
	\begin{subfigure}{.5\textwidth}
		\centering
		\includegraphics[width=.93\linewidth]{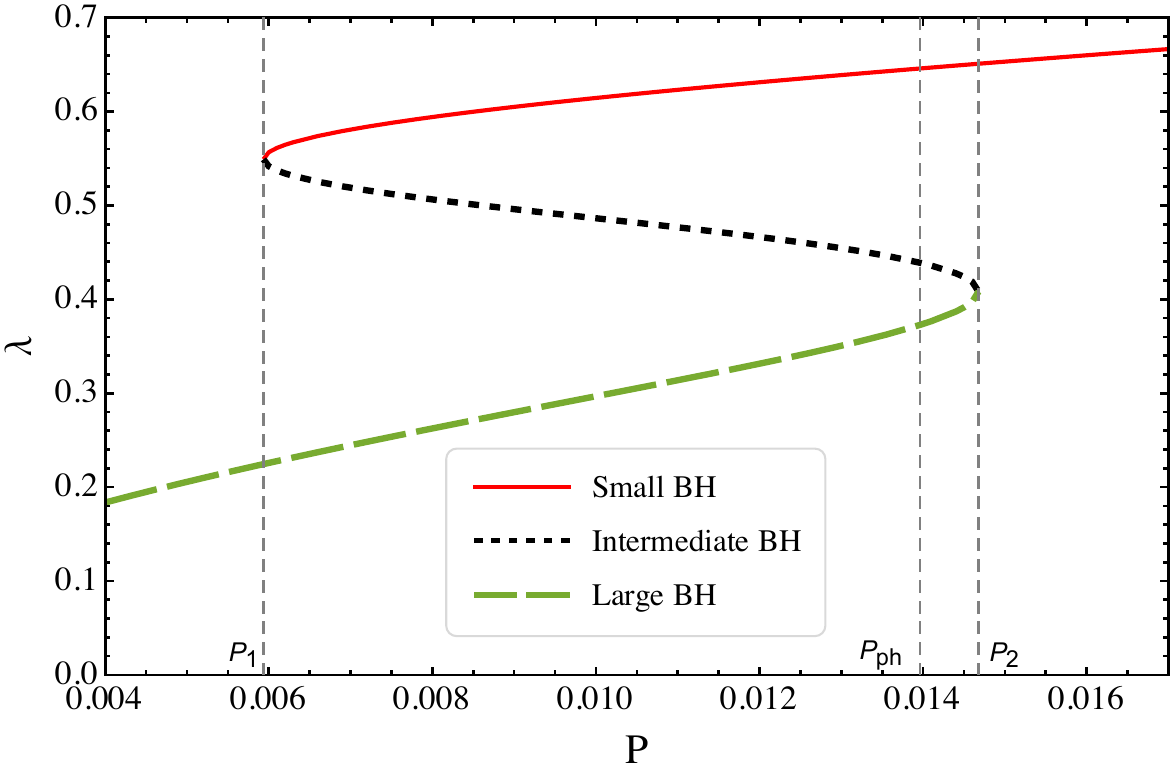}  
		\caption{}
		\label{fig:5DLEm0}
	\end{subfigure}
\begin{subfigure}{.5\textwidth}
	\centering
	\includegraphics[width=.93\linewidth]{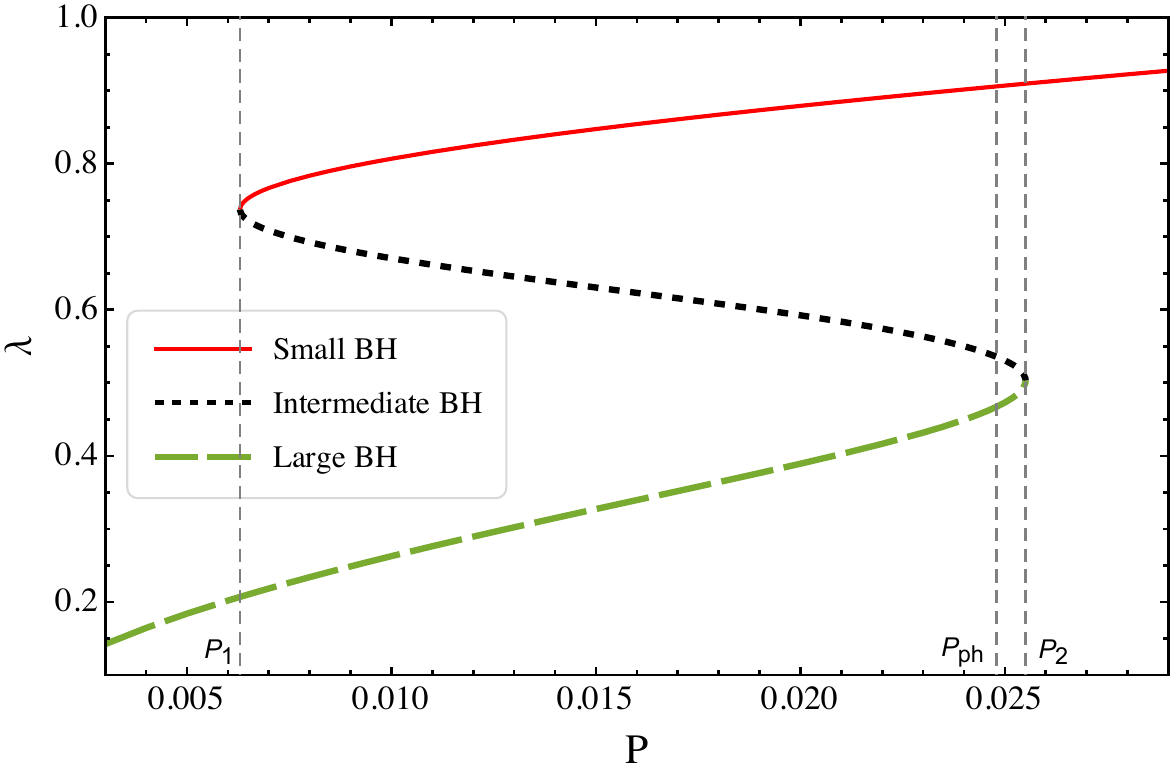}  
	\caption{}
	\label{fig:7DLEm0}
\end{subfigure}
\caption{Lyapunov exponent $\lambda$ of massless particles as a function of the thermodynamic pressure $P$ for five- and seven-dimensional charged regular AdS black holes. The temperature and parameters are set to $T = 0.75 T_{\text{c}}$ and $q=1/2$, $b=4/5$, $L=20$.
(a) five-dimensional case; (b) seven-dimensional case.}
\label{fig:LEm0}
\end{figure}

\begin{figure}[ht]
	\begin{subfigure}{.5\textwidth}
		\centering
		\includegraphics[width=.93\linewidth]{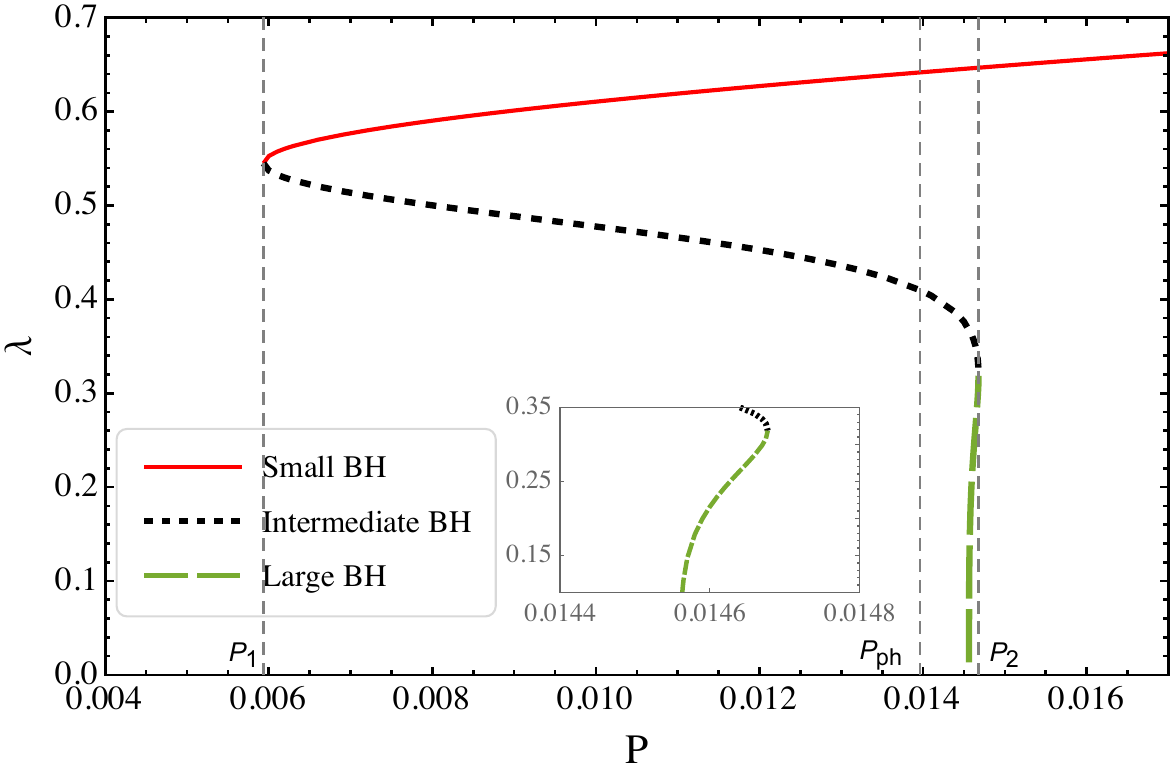}  
		\caption{}
		\label{fig:5DLEm1}
	\end{subfigure}
\begin{subfigure}{.5\textwidth}
	\centering
	\includegraphics[width=.93\linewidth]{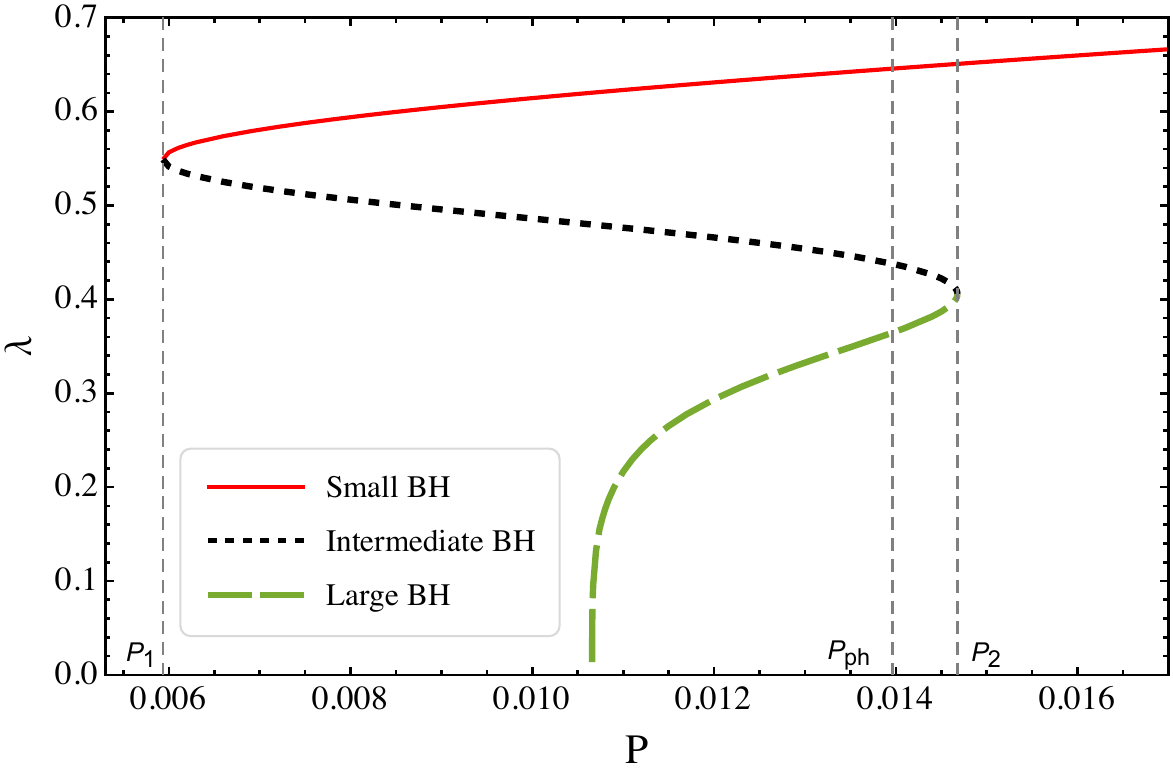}  
	\caption{}
	\label{fig:5DLEm100L}
\end{subfigure}
\caption{Lyapunov exponent $\lambda$ of massive particles as a function of the thermodynamic pressure $P$ for five-dimensional charged regular AdS black holes. The temperature and parameters are set to $T = 0.75 T_{\text{c}}$ and $q=1/2$, $b=4/5$.
(a) L=20; (b) L=100.}
\label{fig:5DLEm1L}
\end{figure}

\begin{figure}[ht]
	\begin{subfigure}{.5\textwidth}
		\centering
		\includegraphics[width=.93\linewidth]{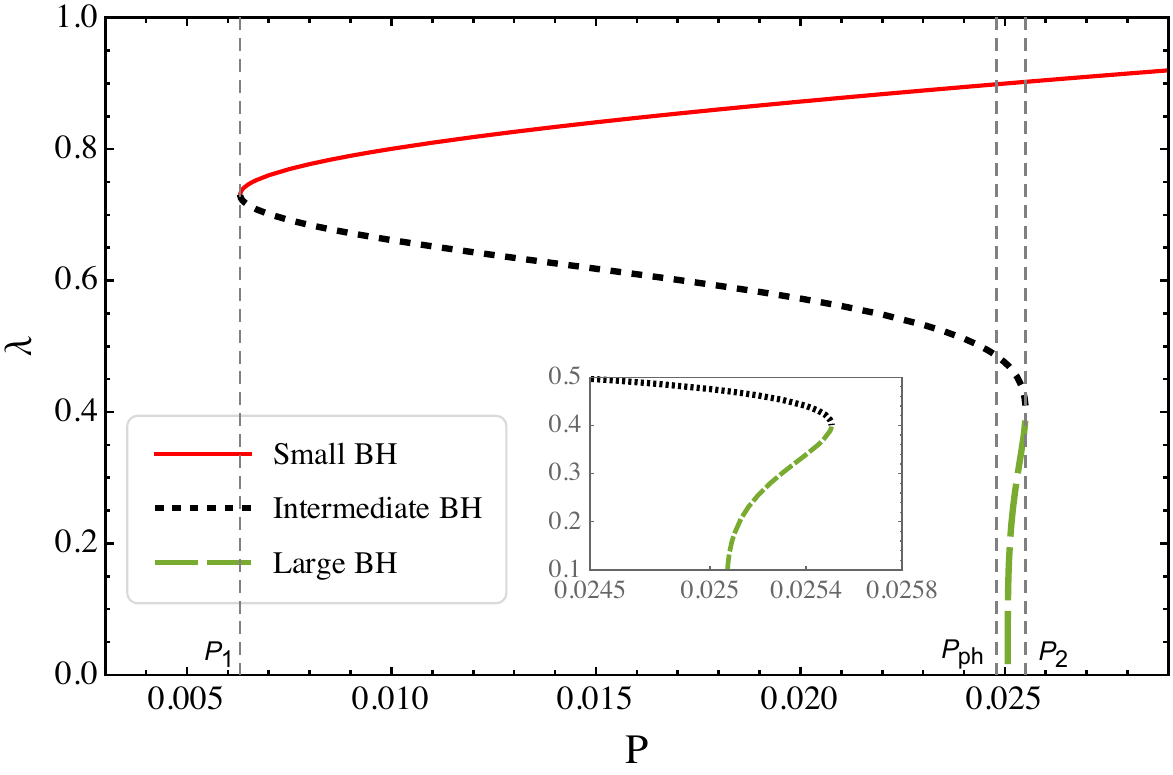}  
		\caption{}
		\label{fig:7DLEm1}
	\end{subfigure}
\begin{subfigure}{.5\textwidth}
	\centering
	\includegraphics[width=.93\linewidth]{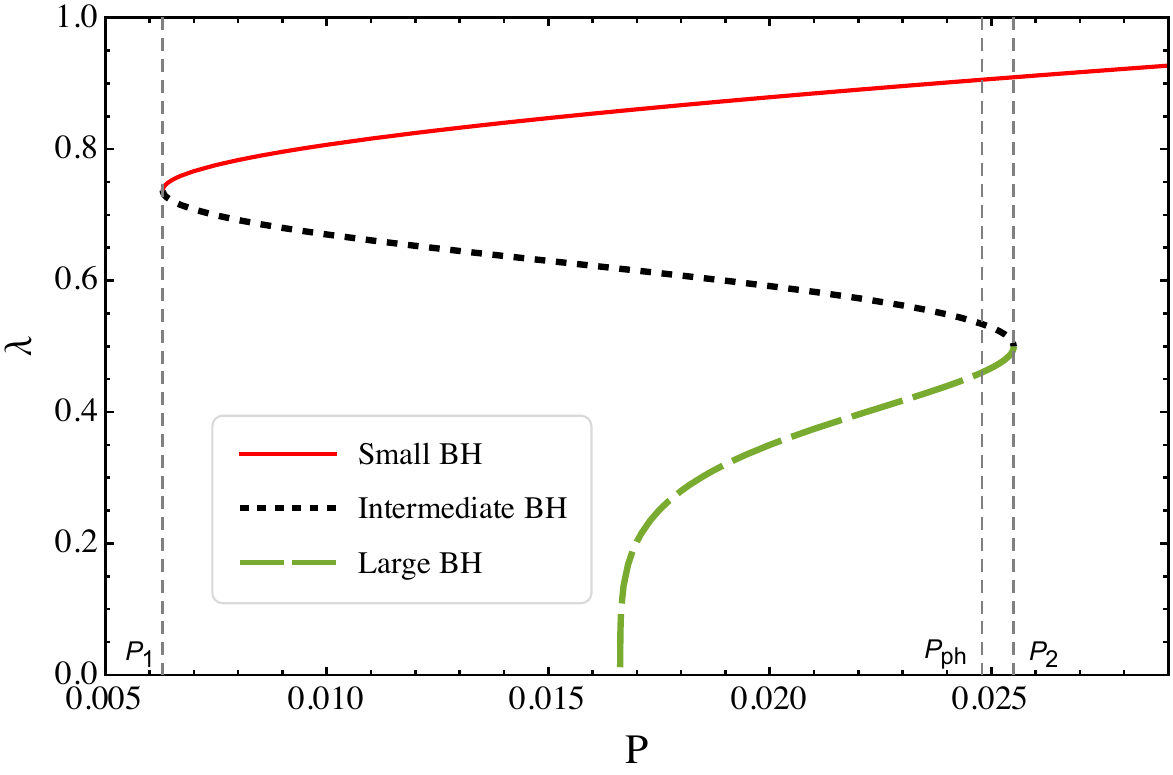}  
	\caption{}
	\label{fig:7DLEm100L}
\end{subfigure}
\caption{Lyapunov exponent $\lambda$ of massive particles as a function of the thermodynamic pressure $P$ for seven-dimensional charged regular AdS black holes. The temperature and parameters are set to $T = 0.75 T_{\text{c}}$ and $q=1/2$, $b=4/5$.
(a) L=20; (b) L=100.}
\label{fig:7DLEm1L}
\end{figure}

We examine the behavior of the Lyapunov exponents for both massless and massive particles as functions of the thermodynamic pressure along the isothermal process at $T = 0.75 T_{\text{c}}$ for five- and seven-dimensional charged regular AdS black holes. The corresponding results are displayed in Figs.~\ref{fig:LEm0},~\ref{fig:5DLEm1L} and~\ref{fig:7DLEm1L}. We find that during the small-large black hole phase transition, the Lyapunov exponent exhibits multivalued behavior similar to that of the $P-v$ curves shown in Figs.~\ref{fig:5DPv} and \ref{fig:7DPv}. Moreover, the Lyapunov exponent in Figs.~\ref{fig:LEm0},~\ref{fig:5DLEm1L} and~\ref{fig:7DLEm1L} displays three distinct branches, corresponding to the small BH, intermediate BH, and large BH phases, respectively. However, it should be noted from Figs.~\ref{fig:5DLEm1} and \ref{fig:7DLEm1} that, for small angular momentum ($L<29.6867$ for five-dimensional black holes and $L<31.9241$ for seven-dimensional black holes), unstable circular orbits for massive particles are absent in the large black hole phase at the phase transition point. In contrast, for large angular momentum ($L>29.6867$ for five-dimensional black holes and $L>31.9241$ for seven-dimensional black holes), such unstable circular orbits emerge in the large black hole phase at the phase transition point, enabling the corresponding Lyapunov exponents to be computed, as shown in Figs.~\ref{fig:5DLEm100L} and~\ref{fig:7DLEm100L}.

Massive particles with small angular momentum can form unstable circular orbits in the small black hole phase, but after the phase transition to the large black hole phase, the same angular momentum may no longer be sufficient to support the formation of such orbits, which prevents one from computing their corresponding Lyapunov exponents. Therefore, probing black hole phase transitions with massive particles’ Lyapunov exponents requires that the particles have sufficiently large angular momentum. In contrast, the existence of unstable circular orbits for massless particles is governed entirely by the spacetime geometry and is independent of the particles’ own properties. This indicates that such unstable circular orbits effectively characterize the intrinsic properties of the background spacetime. We therefore focus, in the following, on the relationship between the Lyapunov exponent of massless particles and the black hole phase transition. For clarity, all references to the Lyapunov exponent in the subsequent discussion pertain to massless particles, unless explicitly stated otherwise.

\begin{table}[h]
	\centering
		\begin{tabular}{p{2em}c || p{2em}c}
					\hline
							\hline
			\multicolumn{2}{c||}{$5D$ AdS black hole} & \multicolumn{2}{c}{$7D$ AdS black hole}\\
			\hline
		 $P / P_{\text{c}}$ &  $\lambda$  &  $P / P_{\text{c}}$ &  $\lambda$  \\
			\hline
			0.46 & 0.340103 & 0.47 & 0.420488 \\
			0.47 & 0.345321 & 0.48 & 0.427720 \\
			0.48 & 0.350741 & 0.49 & 0.435379 \\
			0.49 & 0.356429 & 0.50 & 0.443627 \\
			0.50 & 0.362485 & 0.51 & 0.452747 \\
                0.51 & 0.369076 & 0.52 & 0.463311 \\
			\hline
			0.52 & 0.646846 & 0.53 & 0.907645 \\
			0.53 & 0.648766 & 0.54 & 0.910127 \\
			0.54 & 0.650667 & 0.55 & 0.912590 \\
			0.55 & 0.652551 & 0.56 & 0.915032 \\
                0.56 & 0.654418 & 0.57 & 0.917457 \\
		\hline
			\hline
		\end{tabular}
	\caption{The Lyapunov exponents of massless particles in the five- and seven-dimensional charged regular AdS black holes during an isothermal process at temperature $T=0.75 T_{\text{c}}$. The first column lists the pressure $P$, and the second column shows the corresponding Lyapunov exponents of massless particles. The phase transition point is located at $P=0.515144 P_{\text{c}}$ and $P=0.523072 P_{\text{c}}$ for the five-dimensional and seven-dimensional AdS black holes, respectively.}\label{tab:isotherm}
\end{table}

\begin{table}[h]
	\centering
		\begin{tabular}{p{2em}c || p{2em}c}
					\hline
							\hline
			\multicolumn{2}{c||}{$5D$ AdS black hole} & \multicolumn{2}{c}{$7D$ AdS black hole}\\
			\hline
		 $T / T_{\text{c}}$ &  $\lambda$ & $T / T_{\text{c}}$ &  $\lambda$ \\
			\hline
			0.68 & 0.654568 & 0.68 & 0.929996 \\
			0.69 & 0.653352 & 0.69 & 0.926446 \\
			0.70 & 0.652020 & 0.70 & 0.922698 \\
			0.71 & 0.650558 & 0.71 & 0.918730 \\
			0.72 & 0.648945 & 0.72 & 0.914517 \\
                0.73 & 0.647159 & 0.73 & 0.910029 \\
			\hline
			0.74 & 0.366718 & 0.74 & 0.450333 \\
			0.75 & 0.362485 & 0.75 & 0.443627 \\
			0.76 & 0.359364 & 0.76 & 0.438721 \\
			0.77 & 0.356912 & 0.77 & 0.434861 \\
                0.78 & 0.354910 & 0.78 & 0.431696 \\
		\hline
			\hline
		\end{tabular}
	\caption{The Lyapunov exponents of massless particles in the five- and seven-dimensional charged regular AdS black holes during an isobar process at temperature $P=0.5 P_{\text{c}}$. The first column lists the temperature $T$, and the second column shows the corresponding Lyapunov exponents of massless particles. The phase transition point is located at $T=0.739832 T_{\text{c}}$ and $P=0.734071 P_{\text{c}}$ for the five-dimensional and seven-dimensional AdS black holes, respectively.}\label{tab:isobar}
\end{table}

To further demonstrate that the Lyapunov exponent can serve as a signature of black hole phase transitions, we investigate its behavior across isothermal and isobaric phase transitions. First, we calculate the Lyapunov exponents of test particles along the isothermal process at the temperature $T=0.75 T_{\text{c}}$ on both sides of the phase transition for five- and seven-dimensional black holes. The numerical results are summarized in Table~\ref{tab:isotherm}. As shown in the table, during the isothermal process, the Lyapunov exponent of the large black hole phase gradually increases with the increase of pressure. When the pressure reaches the phase transition pressure, a small-large black hole phase transition occurs, accompanied by a dramatic change in the Lyapunov exponent.

We also calculate the Lyapunov exponents on both sides of the isobaric phase transition at the pressure $P = 0.5 P_{\text{c}}$ for five- and seven-dimensional black holes. The numerical results are listed in Table~\ref{tab:isobar}. For the isobaric process, a similar analysis can be carried out as in the isothermal process. In the table, the Lyapunov exponent of the small black hole decreases with increasing temperature. Upon reaching the small-large black hole coexistence temperature, the black hole undergoes a small-large black hole phase transition, with a significant drop in the Lyapunov exponent.

\subsection{Order parameter and critical exponent}

The observed behavior of the Lyapunov exponent across the phase transition suggests that it may encode valuable information about black hole phase transitions. To explore this possibility, we devote this subsection to a detailed investigation of the evolution of Lyapunov exponents along the coexistence curve during the small–large black hole phase transitions. 

\begin{figure}[ht]
	\begin{subfigure}{.5\textwidth}
		\centering
		\includegraphics[width=.93\linewidth]{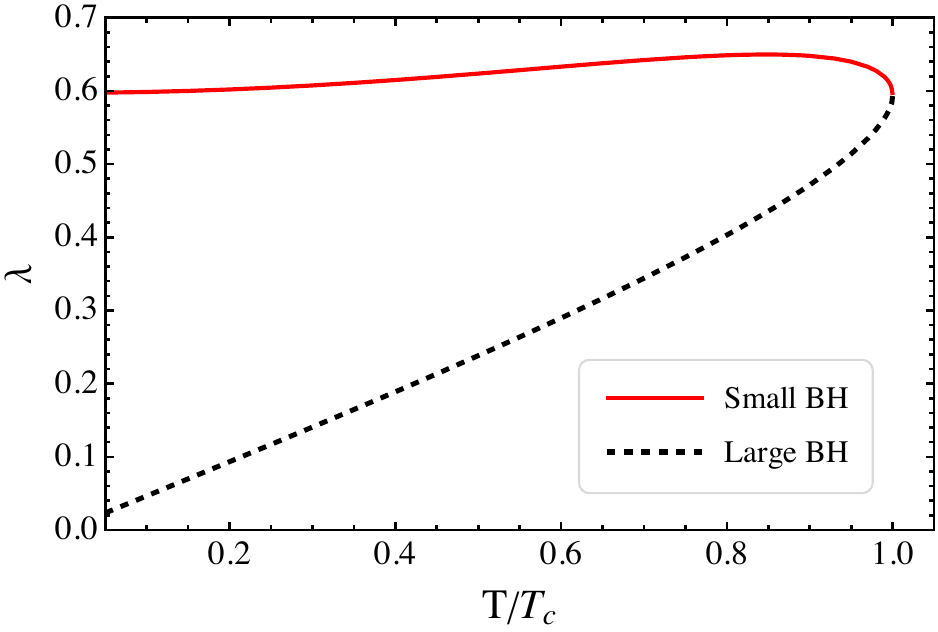}  
		\caption{}
		\label{fig:5DLEsl}
	\end{subfigure}
\begin{subfigure}{.5\textwidth}
	\centering
	\includegraphics[width=.93\linewidth]{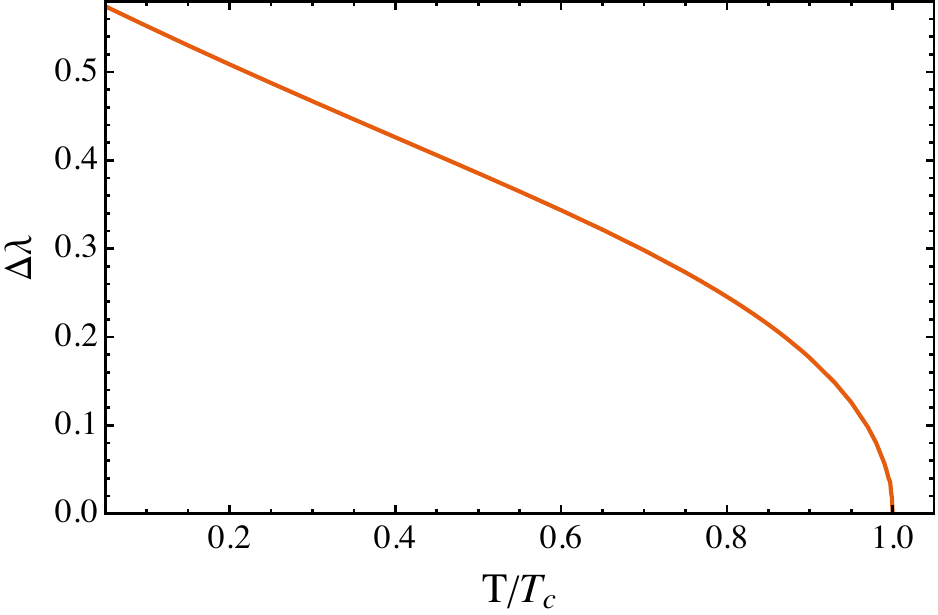}  
	\caption{}
	\label{fig:5DdLE}
\end{subfigure}
\caption{(a) Lyapunov exponents $\lambda$ of the small and large black hole phases as a function of temperature $T$ along the coexistence curve for the five-dimensional charged regular AdS black hole in quasi-topological gravity; (b) The difference in Lyapunov exponents, $\Delta \lambda$, between the small and large black hole phases as a function of temperature $T$ along the coexistence curve for the five-dimensional charged regular AdS black hole. Parameters are chosen as $q=1/2$, $b=4/5~(\alpha = 9/25)$, $L=20$.}
\label{fig:5DdLEsl}
\end{figure}

\begin{figure}[ht]
	\begin{subfigure}{.5\textwidth}
		\centering
		\includegraphics[width=.93\linewidth]{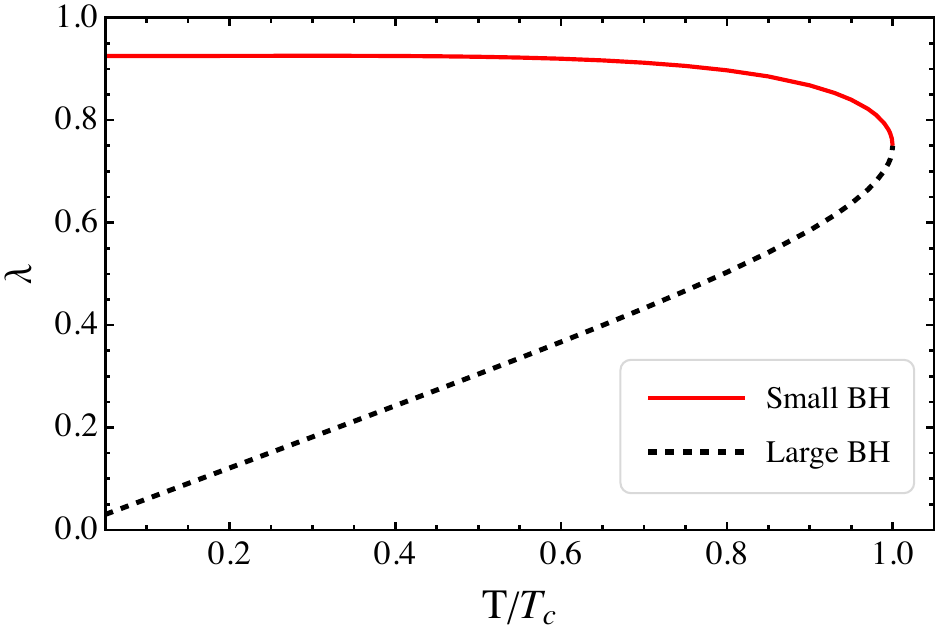}  
		\caption{}
		\label{fig:7DLEsl}
	\end{subfigure}
\begin{subfigure}{.5\textwidth}
	\centering
	\includegraphics[width=.93\linewidth]{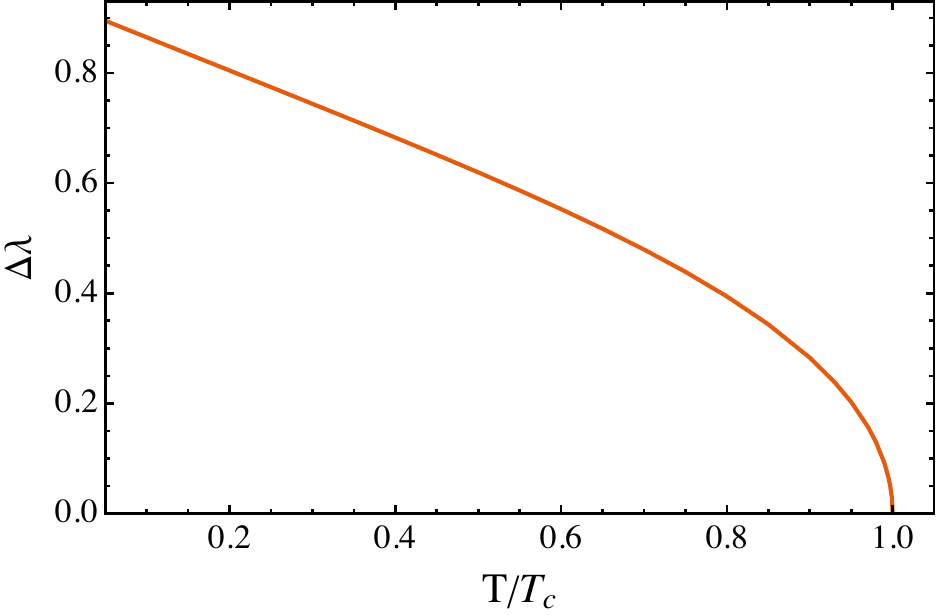}  
	\caption{}
	\label{fig:7DdLE}
\end{subfigure}
\caption{(a). Lyapunov exponents $\lambda$ of the small and large black hole phases as a function of temperature $T$ along the coexistence curve for the seven-dimensional charged regular AdS black hole; (b). The difference in Lyapunov exponents, $\Delta \lambda$, between the small and large black hole phases as a function of temperature $T$ along the coexistence curve for the seven-dimensional charged regular AdS black hole. Parameters are chosen as $q=1/2$, $b=4/5~(\alpha = 1)$, $L=20$.}
\label{fig:7DdLEsl}
\end{figure}

To gain further insight into the dynamical behavior of the black hole during phase transition, we analyze how the Lyapunov exponents of the small and large black hole phases evolve with temperature. 
Along the coexistence curve, the Lyapunov exponents $\lambda$ of the small and large black holes are expressed as functions of the temperature $T$, with the corresponding results for five- and seven-dimensional cases shown in Figs.~\ref{fig:5DLEsl} and \ref{fig:7DLEsl}, respectively. For both cases, the Lyapunov exponents of the small black hole phase increase slowly with temperature, reach a maximum, and then decrease slightly, whereas those of the large black hole phase increase monotonically. At the critical temperature, the Lyapunov exponents of the small and large black hole phases coincide. 

The behavior of the Lyapunov exponents for the small and large black holes indicates that their difference may act as an order parameter characterizing the phase transition. To examine this possibility, we compute the difference in the Lyapunov exponents between the small and large black hole phases along the coexistence curve for both the five- and seven-dimensional black holes. The corresponding results are displayed in Figs.~\ref{fig:5DdLE} and \ref{fig:7DdLE}, respectively. We find that, along the coexistence curve, the difference in Lyapunov exponents between the small and large black hole phases is nonzero and decreases monotonically with increasing temperature. At the critical point, the difference in the Lyapunov exponents becomes zero, exhibiting a behavior that clearly meets the criteria for an order parameter.

To investigate the critical exponent for the order parameter, we expand the Lyapunov exponent near the critical point
\begin{equation}
    \lambda(\rh) = \lambda(r_{\text{hc}}) + \frac{\partial \lambda}{\partial \rh} \Bigg|_{r_{\text{hc}}} ( \rh -r_{\text{hc}} ) + \mathcal{O}(\rh -r_{\text{hc}}),\label{lambdaex}
\end{equation}
where $r_{\text{hc}}$ denotes the horizon radius of the black hole at the critical point. From Eq.~\eqref{lambdaex}, the critical behavior of the difference $\Delta \lambda$ in Lyapunov exponents can be obtained as
\begin{equation}
    \Delta \lambda \propto (r_\text{hl} - r_\text{hs})  \propto (v_\text{hl} - v_\text{hs}), \label{opcb}
\end{equation}
where $r_{\text{hl}}, r_{\text{hs}}$ and $v_{\text{hl}}, v_{\text{hs}}$ denote the horizon radii and specific volumes of the large and small black holes, respectively. From Eq.~\eqref{opcb}, it is clear that the critical behavior of $\Delta \lambda$ is closely related to the specific volume for the small and large black holes. Based on this relation, we focus on the critical behavior of $\Delta v$ for the five-dimensional charged regular AdS black hole in quasi-topological gravity, while the seven-dimensional case can be analyzed using the same approach.

Near the critical point, the equation of state Eq.~(\ref{eos5D}) for the five-dimensional charged regular AdS black hole can be expressed as 
\begin{equation}
   P = \frac{512 q^2}{243 \pi  (1+\phi)^6 v_{\text{c}}^6}-\frac{2 \left[(1+\phi)^4 v_{\text{c}}^4-2 b^4\right]}{3 \pi  \left[(1+\phi)^4 v_{\text{c}}^4-b^4\right]{}^{3/2}}+\frac{(1-t) (1+\phi)^5 T_{\text{c}} v_{\text{c}}^5}{\left[(1+\phi)^4 v_{\text{c}}^4-b^4\right]{}^{3/2}},\label{eoscp}
\end{equation}
where $\phi$ and $t$ represent dimensionless small quantities defined as
\begin{equation}
    \begin{split}
       \phi = \frac{v}{v_{\text{c}}}-1, \quad t = 1-\frac{T}{T_{\text{c}}}.
    \end{split}
\end{equation}
By expanding Eq.~(\ref{eoscp}) in terms of the small quantity $\phi$ and retaining terms up to third order, we obtain:
\begin{equation}
\begin{aligned}
         P&=\frac{v_{\text{c}}^5 (1-t) T_{\text{c}}}{\left(v_{\text{c}}^4-b^4\right){}^{3/2}}-\frac{2 \left(v_{\text{c}}^4-2 b^4\right)}{3 \pi  \left(v_{\text{c}}^4-b^4\right){}^{3/2}}+\frac{512 q^2}{243 \pi  v_{\text{c}}^6}\\
         &- \left[\frac{v_{\text{c}}^5\left(5 b^4+v_{\text{c}}^4\right) (1-t) T_{\text{c}}}{\left(v_{\text{c}}^4-b^4\right){}^{5/2}}-\frac{4 v_{\text{c}}^4 \left(v_{\text{c}}^4-4 b^4\right)}{3 \pi  \left(v_{\text{c}}^4-b^4\right){}^{5/2}}+\frac{1024 q^2}{81 \pi v_{\text{c}}^6}\right] \phi\\
         &+\left[\frac{v_{\text{c}}^5 \left(19 b^4 v_{\text{c}}^4+10 b^8+v_{\text{c}}^8\right) (1-t) T_{\text{c}}}{\left(v_{\text{c}}^4-b^4\right){}^{7/2}}-\frac{2 \left(v_{\text{c}}^{12}-7 b^4 v_{\text{c}}^8-4 b^8 v_{\text{c}}^4\right)}{\pi  \left(v_{\text{c}}^4-b^4\right){}^{7/2}}+\frac{3584 q^2}{81 \pi v_{\text{c}}^6}\right] \phi ^2 \\
         &-\Bigg[\frac{v_{\text{c}}^5  \left(48 b^4 v_{\text{c}}^8+81 b^8 v_{\text{c}}^4+10 b^{12}+v_{\text{c}}^{12}\right) (1-t) T_{\text{c}}}{\left(v_{\text{c}}^4-b^4\right){}^{9/2}}+\frac{28672 q^2}{243 \pi v_{\text{c}}^6}\\ 
         &-\frac{4 \left(2 v_{\text{c}}^{16}-23 b^4 v_{\text{c}}^{12}-45 b^8 v_{\text{c}}^8-4 b^{12} v_{\text{c}}^4\right)}{3 \pi \left(v_{\text{c}}^4-b^4\right){}^{9/2}}\Bigg] \phi ^3+\mathcal{O}\left(\phi ^4\right). \label{Pexpand}
\end{aligned}
\end{equation}
For convenience, Eq.~\eqref{Pexpand} can be recast in the form
\begin{equation}
  P = D(t) + A(t) \phi + B(t) \phi^2 + C(t) \phi^3 +\mathcal{O}(\phi ^4),  
\end{equation}
where $D(t=0) = P_{\text{c}}$. At the phase transition point, by applying Maxwell’s equal-area law, we obtain
\begin{align}
\int_{\phi _s}^{\phi _l} \left[D(t)+A(t) \phi +B(t) \phi ^2+C(t) \phi ^3\right] \, d\phi&= \left(\phi _l-\phi _s\right) \left[D(t)+A(t) \phi _l+B(t) \phi _l^2+C \phi _l^3\right],\\
        D(t)+A(t) \phi_\text{l} +B(t) \phi_\text{l} ^2+C(t) \phi_\text{l} ^3 &= D(t)+A(t) \phi_\text{s} +B(t) \phi_\text{s} ^2+C(t) \phi_\text{s} ^3.
\end{align}
From the above equations, we obtain
\begin{equation}
   \phi_\text{l}-\phi_\text{s} = \frac{2 \sqrt{C(t)^2 \left[B(t)^2-3 A(t) C(t)\right]}}{\sqrt{3} C(t)^2}. \label{phidiff}
\end{equation}
To facilitate the subsequent calculations, it is convenient to express the coefficients $A(t)$, $B(t)$ and $C(t)$ in explicit functional forms. Following Eq.~(\ref{Pexpand}), we take
\begin{equation}
    \begin{aligned}
        A(t) &= a_1 + b_1 t, \\
        B(t) &= a_2 + b_2 t, \\
        C(t) &= a_3 + b_3 t.
    \end{aligned} \label{ABCab}
\end{equation}
Accordingly, Eq.~\eqref{phidiff} takes the following form
\begin{equation}
    \phi_\text{l}-\phi_\text{s} =\frac{2 \sqrt{\sigma(t)}}{\sqrt{3} \left(a_3+b_3 t\right)^2}, \label{philphis}
\end{equation}
where 
\begin{equation}
    \sigma(t)=\left(a_3+b_3 t\right)^2 \left[\left(a_2+b_2 t\right)^2-3 \left(a_1+b_1 t\right) \left(a_3+b_3 t\right)\right].
\end{equation}
Expanding $\sigma(t)$ in a Taylor series around $t=0$, we obtain
\begin{equation}
    \sigma(t) = a_3^2 \left(a_2^2-3 a_1 a_3\right)+ \left(-3 a_3^3 b_1+2 a_2 a_3^2 b_2+2 a_2^2 a_3 b_3-9 a_1 a_3^2 b_3\right) t+\mathcal{O}\left(t^2\right).\label{sigmaphi}
\end{equation}
At the critical point, the values of $\phi_\text{l}$ and $\phi_\text{s}$ coincide. Consequently, we have 
\begin{equation}
    a_3^2 \left(a_2^2-3 a_1 a_3\right)=0.
\end{equation}
It follows that the leading order of $\sigma(t)$ is proportional to $ t$. By expanding the denominator in Eq.~(\ref{philphis}) with respect to $t$, we have
\begin{equation}
    \phi_\text{l} - \phi_\text{s} \propto t^{\frac{1}{2}}.\label{deltaphil}
\end{equation}
Substituting Eq.~\eqref{deltaphil} into Eq.~\eqref{opcb}, we obtain the critical exponent for the order parameter
\begin{equation}
    \Delta \lambda \propto t^{\frac{1}{2}}. \label{deltalmd}
\end{equation}
The same conclusion can be drawn for the seven-dimensional charged regular AdS black hole. The result indicates that the critical exponent associated with the Lyapunov exponent is $1/2$, which coincides with that of the van der Waals fluid~\cite{johnston2014advances} and the Reissner–Nordstr\"om AdS black holes~\cite{Guo:2022kio}.

\section{Discussions and conclusions}\label{sec:con}

In this paper, we investigated the relationship between the thermodynamic phase transitions and the Lyapunov exponents of static, spherically symmetric, charged regular AdS black holes in five- and seven-dimensional quasi-topological gravity. The results demonstrate that the Lyapunov exponents associated with massless particles encode information about black hole phase transitions. First, we investigated the thermodynamic phase transitions of five-dimensional and seven-dimensional charged regular AdS black holes in the extended phase space. We found that, below the critical point, the black holes exhibit van der Waals–like behavior in the $P-v$ phase diagram and swallowtail structure in the Gibbs free energy during the phase transition. At the critical point, the black holes undergo a second-order phase transition; the van der Waals–like behavior and swallowtail behavior disappear. Also, we constructed the coexistence curves for the five- and seven-dimensional charged regular AdS black holes.  

Following the analysis of thermodynamic phase transitions in charged regular AdS black holes, we explored the Lyapunov exponents governing the motion of massless and massive particles. Because the presence of unstable circular orbits for massless particles is dictated solely by the underlying spacetime geometry, rather than by any properties of the particles themselves, such orbits serve as a natural probe of the intrinsic characteristics of the background spacetime. Guided by this observation, our discussion focuses on the Lyapunov exponents associated with these unstable circular null orbits. We find that the Lyapunov exponent exhibits a multivalued behavior with respect to the thermodynamic pressure, together with a dramatic change in the Lyapunov exponent across the phase transition. This suggests that the Lyapunov exponent may indeed encode valuable information about the black hole phase transition. Furthermore, we investigated the behavior of the Lyapunov exponents along the coexistence curve and found that the Lyapunov exponents change discontinuously at the first-order phase transition while changing continuously at the second-order phase transition. At the critical point along the coexistence curve, the difference in Lyapunov exponent $\Delta \lambda$ between large and small black holes exhibits the same critical behavior as the difference in specific volume $\Delta v$, characterized by a critical exponent of $1/2$. This result indicates that the difference of the Lyapunov exponent $\Delta \lambda$ can serve as an order parameter for the black hole phase transition.

The Lyapunov exponent is a measure of a dynamical system's sensitivity to initial conditions, quantifying the average exponential rate at which nearby trajectories diverge or converge over time. Our work further indicates that the Lyapunov exponent might provide a dynamical approach to probing phase transitions in the extended phase space thermodynamics of black holes. Moreover, it reveals the relationship between the instability of black hole circular orbits and black hole phase structure. In addition to the Lyapunov exponent, other strong gravity effects--such as the black hole photon sphere~\cite{Wei:2018aqm,Wei:2017mwc,Yang:2025xck} and quasinormal modes~\cite{Liu:2014gvf, He:2010zb, Jing:2008an}—might also serve as probes of black hole thermodynamic phase transitions and deserve further in-depth investigation.

\acknowledgments

The authors thank Yu-Xiao Liu, Rui-Hua Ni and Shan-Ping Wu for helpful discussions. This work was supported by the National Natural Science Foundation of China (Grants No. 12305065, No. 12475056, and No. 12247101), the China Postdoctoral Science Foundation (Grant No. 2023M731468), the Gansu Province's Top Leading Talent Support Plan, the Fundamental Research Funds for the Central Universities (Grant No. lzujbky-2025-jdzx07), the Natural Science Foundation of Gansu Province (No. 22JR5RA389, No.25JRRA799), and the `111 Center' under Grant No. B20063.
\appendix 
\renewcommand{\thesection}{\Alph{section}} 
\renewcommand{\theequation}{\thesection.\arabic{equation}} 
\setcounter{equation}{0} 

\section{Lyapunov exponent} \label{appendixA} 

An N-dimensional dynamical system $\mathbf{X}=(x^1, x^2, ..., x^N)^T$ can be described by first-order differential equations as follows:
\begin{equation}
    \frac{dx^i}{d t} = F^i (\mathbf{x}). 
\end{equation}
A variational analysis of the above system yields
\begin{equation}
    \frac{d \delta x^i (t)}{d t} = K^i_{\;j} (t) \delta x^j (t),\label{ddkx}
\end{equation}
where
\begin{equation}
    K^i_{\;j} (t) = \frac{\partial F^i (\mathbf{x})}{\partial x^j} \Big|_{\mathbf{x}(t)}.
\end{equation}
The solution to Eq.~(\ref{ddkx}) can be expressed as
\begin{equation}
    \delta x^i (t) = U^i_{\;j} (t, 0) \delta x^j(0),
\end{equation}
where $U^i_{\;j} (t, 0)$ is the evolution matrix satisfying the following equation:
\begin{equation}
    \frac{d U^i_{\;j} (t, 0)}{d t} = K^i_{\;k} (t) U^k_{\;j} (t, 0).
\end{equation}

A dynamical system can be described by an orbit $\mathbf{X}(t)$ in an 
$n$-dimensional phase space. Consider a nearby trajectory $\mathbf{X}(t) + \delta \mathbf{X}(t)$. The maximal Lyapunov exponent associated with the reference orbit $\mathbf{X}(t)$ is defined as
\begin{equation}
    \lambda = \lim_{t \to \infty} \lim_{|\delta  \mathbf{X}(0)| \to 0} \frac{1}{t} \log \frac{|\delta  \mathbf{X}(t)|}{|\delta  \mathbf{X}(0)|}, \label{Lyapunov}
\end{equation}
where $\delta  \mathbf{X} (t)$ denotes the perturbation vector along the direction of maximal growth, corresponding to the most unstable mode.

In this study, we focus on the dynamical system describing the orbital motion of particles outside the black hole horizon. The corresponding phase space is two-dimensional and can be represented by the state vector
\begin{equation}
    \mathbf{X}=
\begin{pmatrix}
r \\
p_{r}
\end{pmatrix},
\end{equation}
where $p_r$ denotes the canonical momentum conjugate to the radial coordinate $r$.

Using Eq.~(\ref{particleL}), the Hamiltonian describing the motion of particles outside the black hole horizon can be obtained as
\begin{equation}
    \begin{aligned}
        \mathcal{H} &= p_t \dot{t} + p_r \dot{r} + p_{\varphi} \dot{\varphi} - \mathcal{L} \\
                    &=\frac{V_{\text{eff}}(r)-E^2}{2 f(r)} + \frac{f(r) {p_r}^2}{2}-\frac{\delta}{2},
    \end{aligned}
\end{equation}
where $p_t, p_r$ and $p_{\varphi}$ are the canonical momenta conjugate to the coordinates $t, r$ and $\varphi$, respectively. Utilizing Hamilton's equations, one can derive
\begin{equation}
    \begin{aligned}
        \dot{r} &= \frac{\partial \mathcal{H}}{\partial p_r} = f(r) p_r, \\
      \dot{p}_r &= -\frac{\partial \mathcal{H}}{r} \\
                &=-\frac{V'_{\text{eff}}(r)}{2 f(r)} + \frac{(V_{\text{eff}}(r) - E^2)f'(r)}{2 f(r)^2} - \frac{f'(r) p_r^2}{2}.
    \end{aligned} \label{dotrpr}
\end{equation}
Considering a small perturbation around the circular orbit located at $r=r_{\text{c}}$, and under the conditions $V'_{\text{eff}}(r_{\text{c}}) = 0$ and $ V_{\text{eff}}(r_{\text{c}}) - E^2 =0 $, neglecting higher-order infinitesimal terms, the linearized equations governing the perturbations can be written as
\begin{equation}
    \begin{aligned}
        \delta \dot{r} &= \dot{t} \frac{d \delta r}{d t},\\
        \delta \dot{p}_r &= \dot{t} \frac{d \delta p_r}{d t}.
    \end{aligned} \label{drdpr}
\end{equation}
Using Eqs.~(\ref{particleL}),~(\ref{dotrpr}) and~(\ref{drdpr}), we can obtain
\begin{equation}
    \begin{pmatrix}
        \frac{d \delta x^1}{dt} \\
        \frac{d \delta x^2}{dt}
    \end{pmatrix}
    =
    \begin{pmatrix}
        \frac{d \delta r}{dt} \\
        \frac{d \delta p_r}{dt}
    \end{pmatrix}
    =
    \mathbf{K} 
    \begin{pmatrix}
         \delta r \\
        \delta p_r
    \end{pmatrix}
    =\mathbf{K} 
    \begin{pmatrix}
         \delta x^1 \\
        \delta x^2
    \end{pmatrix}, \label{kxrpr}
\end{equation}
where 
\begin{equation}
    \mathbf{K} = 
    \begin{pmatrix}
        0 & K^1 \\
        K^2& 0
    \end{pmatrix}
    =
    \begin{pmatrix}
        0 & \dot{t}^{-1} f(r_{\text{c}}) \\
        -\dot{t}^{-1} \frac{V''_{\text{eff}} (r_{\text{c}})}{2f(r_{\text{c}})} & 0.
    \end{pmatrix}. \label{kkk}
\end{equation}
To identify the direction in which $\delta \mathbf{X}$ exhibits the most rapid variation, we diagonalize the matrix $\mathbf{K}$ and perform a corresponding orthogonal transformation on the $\delta \mathbf{X}$. This yields
\begin{equation}
    \begin{pmatrix}
        0 & K^1 \\
        K^2& 0
    \end{pmatrix}
    = \mathbf{V}^{-1}
    \begin{pmatrix}
        +\sqrt{K^1 K^2} & 0 \\
        0 & -\sqrt{K^1 K^2}
    \end{pmatrix} 
    \mathbf{V},
\end{equation}
\begin{equation}
    \begin{pmatrix}
        \delta y^1(t) \\
        \delta y^2(t)
    \end{pmatrix}
    = \mathbf{V}
    \begin{pmatrix}
        \delta x^1(t) \\
        \delta x^2(t)
    \end{pmatrix}.
\end{equation}
Accordingly, Eq.~(\ref{kxrpr}) can be reformulated as
\begin{equation}
    \begin{pmatrix}
        \frac{d \delta y^1(t)}{dt} \\
        \frac{d \delta y^2(t)}{dt}
    \end{pmatrix}
    =
    \begin{pmatrix}
        +\sqrt{K^1 K^2} & 0 \\
        0 & -\sqrt{K^1 K^2}
    \end{pmatrix} 
    \begin{pmatrix}
        \delta y^1(t) \\
        \delta y^2(t)
    \end{pmatrix}.    
\end{equation}
From the above results, it follows that the direction along which $\delta \mathbf{X}$ varies most rapidly corresponds to the $\delta y^1(t)$. Moreover, $\delta y^1(t)$ satisfies the equation
\begin{equation}
    \delta y^1(t) = e^{\sqrt{K^1 K^2} t} \delta y^1(0).
\end{equation}
By combining this result with the definition of the Lyapunov exponent Eq.~(\ref{Lyapunov}),
the Lyapunov exponent associated with the circular orbit can be expressed as
\begin{equation}
    \lambda = \sqrt{K^1 K^2}.
\end{equation}

Employing Eq.~\eqref{kkk} together with the conditions $E = -f(r_{\text{c}}) \dot{t}$ and $E^2 = V_{\text{eff}} (r_{\text{c}})$, the Lyapunov exponent corresponding to the circular orbit of a massless particle can be expressed as
\begin{equation}
    \lambda = \sqrt{-\frac{f(r_{\text{c}}) r_{\text{c}} ^2}{2 L^2} V''_{\text{eff}}(r_{\text{c}})}.
\end{equation}
Similarly, using Eq.~\eqref{kkk} and the conditions $E = -f(r_{\text{c}}) \dot{t}$, $E^2 = V_{\text{eff}} (r_{\text{c}})$, and $V'_{\text{eff}}(r_{\text{c}}) = 0$, the Lyapunov exponent associated with the circular orbit of a massive particle is obtained as
\begin{equation}
    \lambda = \frac{1}{2} \sqrt{[r_{\text{c}} f'(r_{\text{c}}) - 2 f(r_{\text{c}})] V''_{\text{eff}} (r_{\text{c}})}.
\end{equation}

\providecommand{\href}[2]{#2}\begingroup\raggedright\endgroup
\end{document}